\documentclass[screen,acmsmall]{acmart}
\usepackage{mathpartir}
\usepackage[utf8]{inputenc}
\usepackage{fontenc}
\usepackage{microtype}
\usepackage{listings,array,multirow,wrapfig,xspace,booktabs,subcaption}
\usepackage{xcolor,tikz, graphicx}
\usetikzlibrary{positioning}
\graphicspath{ {graphs/} }

\newcommand{\GNUR} {{\small{GNU\,R}}\xspace}
\newcommand{\GNURVERSION} {3.5.0\xspace}

\newcommand{\rcode}[1] {\texttt{#1}}
\newcommand{\EXP}{\ensuremath{\mathtt{e}}\xspace}
\newcommand{\EXPp}{\ensuremath{\mathtt{e'}}\xspace}
\newcommand{\EXPpp}{\ensuremath{\mathtt{e''}}\xspace}

\newcommand{\VAL}{\ensuremath{v}\xspace}
\newcommand{\VAR}{\ensuremath{\mathtt{x}}\xspace}

\newcommand{\STR}{\ensuremath{\mathtt{s}}\xspace}

\newcommand{\REF}{\ensuremath{l}\xspace}
\newcommand{\NULLREF}{\ensuremath{\mathtt{\bot}}\xspace}
\newcommand{\REFp}{\ensuremath{l'}\xspace}
\newcommand{\REFpp}{\ensuremath{l''}\xspace}

\newcommand{\CTXT}{\ensuremath{\mathbb C}\xspace}

\newcommand{\FLAG}{\ensuremath{R}\xspace}
\newcommand{\SEEN}{\ensuremath{\uparrow}\xspace}
\newcommand{\UNSEEN}{\ensuremath{\downarrow}\xspace}
\newcommand{\HEAP}{\ensuremath{H}\xspace}
\newcommand{\HEAPp}{\ensuremath{H'}\xspace}
\newcommand{\HEAPpp}{\ensuremath{H''}\xspace}
\newcommand{\EMPTYHEAP}{\ensuremath{\epsilon}\xspace}
\newcommand{\FRAME}{\ensuremath{F}\xspace}
\newcommand{\FRAMEp}{\ensuremath{F'}\xspace}

\newcommand{\STACK}{\ensuremath{S}\xspace}
\newcommand{\EMPTYSTACK}{\ensuremath{\epsilon}\xspace}
\newcommand{\ENV}{\ensuremath{E}\xspace}
\newcommand{\ENVp}{\ensuremath{E'}\xspace}
\newcommand{\ENVpp}{\ensuremath{E''}\xspace}
\newcommand{\EMPTYENV}{\ensuremath{\epsilon}\xspace}
\renewcommand{\C}[1]{\ensuremath{\CTXT[#1]}}
\def\OR{\:\mid\:}
\newcommand{\sglprime}[1]{\ensuremath{#1^\prime}}

\newcommand{\seqtwo}[2]{\ensuremath{#1 \boldsymbol{\cdot} #2}}
\newcommand{\seqthree}[3]{\ensuremath{\seqtwo{#1}{\seqtwo{#2}{#3}}}}
\newcommand{\config}[2]{\ensuremath{#1; \, #2}}
\newcommand{\evalrel}[3]{\ensuremath{#1 \, #2 \, #3}}
\newcommand{\dblrel}[4]{\evalrel{\config{#1}{#2}}{\rightarrow}{\config{#3}{#4}}}

\newcommand{\singlet}[1]{\ensuremath{(#1)}}
\newcommand{\pair}[2]{\ensuremath{(#1, \, #2)}}
\newcommand{\triplet}[3]{\ensuremath{(#1, \, #2, \, #3)}}
\newcommand{\quadruplet}[4]{\ensuremath{(#1, \, #2, \, #3, \, #4)}}

\newcommand{\fresh}[1]{\ensuremath{\mathit{fresh}~#1}}
\newcommand{\extend}[3]{\ensuremath{#1[#2 \mapsto #3]}}

\newcommand{\funcall}[2]{\ensuremath{\mathit{#1} #2}\xspace}

\newcommand{\callone}[2]{\funcall{#1}{\singlet{#2}}}
\newcommand{\calltwo}[3]{\funcall{#1}{\pair{#2}{#3}}}
\newcommand{\callthree}[4]{\funcall{#1}{\triplet{#2}{#3}{#4}}}

\newcommand{\env}{\ensuremath{\mathtt{env}}}
\newcommand{\subst}[1]{\callone{\mathtt{subst}}{#1}}

\newcommand{\str}[1]{\callone{string}{#1}}

\newcommand{\domain}[1]{\callone{dom}{#1}}
\newcommand{\eval}[2]{\calltwo{\mathtt{eval}}{#1}{#2}}
\newcommand{\delay}[3]{\callthree{\mathtt{delay}}{#1}{#2}{#3}}

\newcommand{\get}[3]{\callthree{get}{#1}{#2}{#3}}

\newcommand{\assign}[2]{\ensuremath{#1 \leftarrow #2}}

\newcommand{\concat}[2]{\ensuremath{#1 \, \mathtt{\#} \, #2}}
\newcommand{\closure}[4]{\ensuremath{\pair{\lambda{#1\!=\!#2}.#3}{#4}}\xspace}
\newcommand{\function}[3]{\ensuremath{\mathtt{fun}(#1\!=\!#2)\,#3}\xspace}
\newcommand{\envval}[1]{\ensuremath{\mathtt{env}(#1)}\xspace}
\newcommand{\promval}[1]{\ensuremath{\mathtt{prom}(#1)}\xspace}

\newcolumntype{L}{>{$}l<{$}} 

\let\OldLabTirName \LabTirName
\newcommand {\LabTirName}[1] {\OldLabTirName{\large{[#1]}}}
\newcommand {\irlabel}[1] {\ensuremath{\mathtt{#1}}}

\newcommand{\BUILTINCOUNT} {680\,\xspace}
\newcommand{\SPECIALCOUNT} {46\,\xspace}

\newcommand {\CountCorpusPackages} {16,707\xspace} 
\newcommand {\ProgramCount} {232,290\xspace}
\newcommand {\CntCranPkg} {14,762\xspace}
\newcommand {\CntBiocSoftPkg} {1,741\xspace}
\newcommand {\CntBiocDataPkg} {1,319\xspace}
\newcommand {\CntBiocWflPkg} {27\xspace}
\newcommand {\CntBiocPkg} {3,087\xspace}
\newcommand {\CntTotalPkg} {17,849\xspace}
\newcommand {\CntInstPkg} {17,479\xspace}
\newcommand {\CountInstalledExamplesRScripts} {220\,K\xspace} 
\newcommand {\CountTracedExamplesRScripts} {202\,K\xspace} 
\newcommand {\CountInstalledTestsRScripts} {44.1\,K\xspace} 
\newcommand {\CountTracedTestsRScripts} {23.3\,K\xspace} 
\newcommand {\CountInstalledVignettesRScripts} {9.8\,K\xspace} 
\newcommand {\CountTracedVignettesRScripts} {6.6\,K\xspace} 
\newcommand {\CountInstalledExamplesRCode} {1.6\,M\xspace} 
\newcommand {\CountTracedExamplesRCode} {1.6\,M\xspace} 
\newcommand {\CountInstalledTestsRCode} {2.7\,M\xspace} 
\newcommand {\CountTracedTestsRCode} {1.3\,M\xspace} 
\newcommand {\CountInstalledVignettesRCode} {614\,K\xspace} 
\newcommand {\CountTracedVignettesRCode} {327\,K\xspace} 
\newcommand {\CountTracedSourceRCode} {25.6\,M\xspace} 
\newcommand {\CountTracedSourceCCode} {10.4\,M\xspace} 
 
\newcommand{\CountRDyntraceCodeDiff} {1,886} 
\newcommand{\CountGnuRLoc} {542,809} 
\newcommand {\CountEnvirLoc} {2,864\xspace}
\newcommand {\CountEnvirFunDef} {131\xspace}

\newcommand {\SizeTracerRawData} {5.1 TB\xspace} 
\newcommand {\SizeReducedData} {76 GB\xspace} 
\newcommand {\SizeTotalTotal} {5.2 TB\xspace} 

\newcommand {\CountTracingPipelineCorpusPackage} {16,707\xspace}
\newcommand {\CountTracingPipelineCorpusScript} {232,290\xspace}
\newcommand {\TimeTracingPipelineCorpus} {10h\xspace} 

\newcommand {\CountTracingPipelineTraceFile} {2.8 M\xspace}
\newcommand {\SizeTracingPipelineTraceFile} {5.1 TB\xspace}
\newcommand {\TimeTracingPipelineTrace} {2d 4h\xspace} 

\newcommand {\CountTracingPipelineReduceFile} {7.4 M\xspace}
\newcommand {\SizeTracingPipelineReduceFile} {76 GB\xspace}
\newcommand {\TimeTracingPipelineReduce} {1d 15h\xspace} 

\newcommand {\CountTracingPipelineCombineFile} {842 \xspace}
\newcommand {\SizeTracingPipelineCombineFile} {20.4 GB\xspace}
\newcommand {\TimeTracingPipelineCombine} {3h 30m\xspace} 

\newcommand {\CountTracingPipelineMergeFile} {36 \xspace}
\newcommand {\SizeTracingPipelineMergeFile} {21 GB\xspace}
\newcommand {\TimeTracingPipelineMerge} {1h 24m\xspace} 

\newcommand {\CountTracingPipelineSummarizeFile} {77 \xspace}
\newcommand {\SizeTracingPipelineSummarizeFile} {1.5 GB\xspace}
\newcommand {\TimeTracingPipelineSummarize} {2h 18m\xspace} 

\newcommand {\CountTracingPipelineReportGraph} {15 \xspace}
\newcommand {\CountTracingPipelineReportMacro} {268 \xspace}
\newcommand {\TimeTracingPipelineReport} {37 s\xspace}
\newcommand {\CountCall} {831\,B\xspace} 
\newcommand {\CountFunction} {698.4\,K\xspace} 
\newcommand {\CountJumpedCall} {23.7\,B\xspace} 
\newcommand {\PercBuiltinCall} {26.7\%\xspace} 
\newcommand {\PercSpecialCall} {60.3\%\xspace} 
\newcommand {\CountBuiltinFunction} {157\xspace} 
\newcommand {\CountSpecialFunction} {33\xspace} 
\newcommand {\CountThreeFunction} {34.1\,K\xspace} 
\newcommand {\PercThreeCall} {2\%\xspace} 
\newcommand {\CountFourFunction} {60.2\,K\xspace} 
\newcommand {\PercFourCall} {0.8\%\xspace} 
\newcommand {\CountOrdFunction} {415.3\,K\xspace} 
\newcommand {\CountDelayedAssignCall} {82\,M\xspace} 
\newcommand {\CountForceCall} {101.8\,M\xspace} 
\newcommand {\CountForceAndCallCall} {1.3\,B\xspace} 
\newcommand {\CountSubstituteCall} {1.7\,B\xspace} 
\newcommand {\PercSubstituteCall} {2\%\xspace} 
\newcommand {\CountParameter} {2.1\,M\xspace} 
\newcommand {\CountArgument} {261\,B\xspace} 
\newcommand {\PercDefArgument} {25.1\%\xspace} 
\newcommand {\PercNonDefArgument} {72.1\%\xspace} 
\newcommand {\PercMissingArgument} {2.8\%\xspace} 
\newcommand {\CountPromiseArgument} {261\,B\xspace} 
\newcommand {\CountTotEventRate} {803.1\,K\xspace} 
\newcommand {\CountExpression} {1.7\,T\xspace} 
\newcommand {\PercSingleFunctionPackages} {7.0\%\xspace} 
\newcommand {\PercManyFunctionPackages} {59.2\%\xspace} 
\newcommand {\PercSingleCallClosures} {18.9\%\xspace} 
\newcommand {\PercManyCallClosures} {45.0\%\xspace} 
\newcommand {\PercSingleParameterClosures} {43.7\%\xspace} 
\newcommand {\PercManyParameterClosures} { 9.8\%\xspace} 
\newcommand {\CodeMaximumParameterFunctionName} {\code{meta::forest.meta}\xspace} 
\newcommand {\CountMaximumParameterCount} {199\xspace} 
\newcommand {\CountProm} {270.9\,B\xspace} 
\newcommand {\PercArgProm} {94.3\%\xspace} 
\newcommand {\PercForceProm} {87.3\%\xspace} 
\newcommand {\PercDispatch} {1\%\xspace} 
\newcommand {\PercArgumentPromiseExpressionFunctionCall} {17.3\%\xspace} 
\newcommand {\PercStrictFun} {83.7\%\xspace} 
\newcommand {\CountFun} {388.3\,K\xspace} 
\newcommand {\CountStrPackLess} {2.6\,K\xspace} 
\newcommand {\CountStrFunLess} {81.7\,K\xspace} 
\newcommand {\PercStrPackLess} {16.9\%\xspace} 
\newcommand {\PercStrFunLess} {21.0\%\xspace} 
\newcommand {\PercAlways} {87.6\%\xspace} 
\newcommand {\PercNever} {6.4\%\xspace} 
\newcommand {\PercSometimes} {6.0\%\xspace} 
\newcommand {\PercTwo} { 9.6\%\xspace} 
\newcommand {\PercThree} {3\%\xspace} 
\newcommand {\CountEscapeArgument} {11.6\,M\xspace} 
\newcommand {\PercLookMeta} {0.2\%\xspace} 
\newcommand {\PercMetaOnly} {0.5\%\xspace} 
\newcommand {\CountPackMeta} {2\,K\xspace} 
\newcommand {\PercPackMeta} {11.9\%\xspace} 
\newcommand {\CountEventAlloc} {2.4\,T\xspace} 
\newcommand {\CountEventArgListEnter} {111\,B\xspace} 
\newcommand {\CountEventArgListExit} {111\,B\xspace} 
\newcommand {\CountEventCall} {831\,B\xspace} 
\newcommand {\CountEventCtxtEnter} {143\,B\xspace} 
\newcommand {\CountEventCtxtExit} {143\,B\xspace} 
\newcommand {\CountEventCtxtJmp} {6\,B\xspace} 
\newcommand {\CountEventCtxtSensEnter} {78\,B\xspace} 
\newcommand {\CountEventCtxtSensExit} {78\,B\xspace} 
\newcommand {\CountEventDeserialize} {109\,B\xspace} 
\newcommand {\CountEventEval} {1.7\,T\xspace} 
\newcommand {\CountEventFree} {2.3\,T\xspace} 
\newcommand {\CountEventGC} {4.8\,M\xspace} 
\newcommand {\CountEventPromEnter} {223\,B\xspace} 
\newcommand {\CountEventPromExit} {223\,B\xspace} 
\newcommand {\CountEventPromRead} {102\,B\xspace} 
\newcommand {\CountEventPromSubst} {1.9\,B\xspace} 
\newcommand {\CountEventReturn} {807.4\,B\xspace} 
\newcommand {\CountEventSThreeDispatch} {2.6\,B\xspace} 
\newcommand {\CountEventSFourDispatch} {936\,M\xspace} 
\newcommand {\CountEventSubstitute} {1.7\,B\xspace} 
\newcommand {\CountEventVarDef} {365\,B\xspace} 
\newcommand {\CountEventVarRead} {1.6\,T\xspace} 
\newcommand {\CountEventVarRem} {8.4\,B\xspace} 
\newcommand {\CountEventVarWrite} {140\,B\xspace} 
\newcommand {\PercCallerEnvSubstituteCalls} {99\%\xspace} 
\newcommand {\PercNewEnvSubstituteCalls} {0.7\%\xspace} 
\newcommand {\CountDynamicEnvSubstituteCalls} {209.9\,K\xspace} 
\newcommand {\CountLifecycleTotal} {28.1\,K\xspace} 
\newcommand {\CountLifecycleUnescapedArguments} {19.2\,K\xspace} 
\newcommand {\CountLifecycleEscapedArguments} {7.3\,K\xspace} 
\newcommand {\CountLifecycleNonArguments} {6.2\,K\xspace} 
\newcommand {\TextLifecycleEscapedArgumentsa} {\code{EF}\xspace} 
\newcommand {\PercLifecycleEscapedArgumentsa} {70.7\%\xspace} 
\newcommand {\TextLifecycleEscapedArgumentsb} {\code{FRER}\xspace} 
\newcommand {\PercLifecycleEscapedArgumentsb} {5.4\%\xspace} 
\newcommand {\TextLifecycleEscapedArgumentsc} {\code{FER}\xspace} 
\newcommand {\PercLifecycleEscapedArgumentsc} {5\%\xspace} 
\newcommand {\TextLifecycleEscapedArgumentsd} {\code{FRRER}\xspace} 
\newcommand {\PercLifecycleEscapedArgumentsd} {4\%\xspace} 
\newcommand {\TextLifecycleEscapedArgumentse} {\code{EFR}\xspace} 
\newcommand {\PercLifecycleEscapedArgumentse} {3\%\xspace} 
\newcommand {\TextLifecycleEscapedArgumentsf} {\code{FRRERR}\xspace} 
\newcommand {\PercLifecycleEscapedArgumentsf} {1\%\xspace} 
\newcommand {\TextLifecycleEscapedArgumentsg} {\code{EFRR}\xspace} 
\newcommand {\PercLifecycleEscapedArgumentsg} {1\%\xspace} 
\newcommand {\TextLifecycleUnescapedArgumentsa} {\code{F}\xspace} 
\newcommand {\PercLifecycleUnescapedArgumentsa} {70.5\%\xspace} 
\newcommand {\PercLifecycleUnescapedArgumentsb} {11.9\%\xspace} 
\newcommand {\TextLifecycleUnescapedArgumentsc} {\code{FR}\xspace} 
\newcommand {\PercLifecycleUnescapedArgumentsc} { 9.5\%\xspace} 
\newcommand {\TextLifecycleUnescapedArgumentsd} {\code{FRR}\xspace} 
\newcommand {\PercLifecycleUnescapedArgumentsd} {3\%\xspace} 
\newcommand {\TextLifecycleUnescapedArgumentse} {\code{FRRRR}\xspace} 
\newcommand {\PercLifecycleUnescapedArgumentse} {2\%\xspace} 
\newcommand {\TextLifecycleUnescapedArgumentsf} {\code{FRRR}\xspace} 
\newcommand {\PercLifecycleUnescapedArgumentsf} {1\%\xspace} 
\newcommand {\TextLifecycleUnescapedArgumentsg} {\code{M}\xspace} 
\newcommand {\PercLifecycleUnescapedArgumentsg} {0.5\%\xspace} 
\newcommand {\TextLifecycleNonArgumentsa} {\code{A}\xspace} 
\newcommand {\PercLifecycleNonArgumentsa} {67.5\%\xspace} 
\newcommand {\PercLifecycleNonArgumentsb} {16.5\%\xspace} 
\newcommand {\TextLifecycleNonArgumentsc} {\code{AR}\xspace} 
\newcommand {\PercLifecycleNonArgumentsc} {11.2\%\xspace} 
\newcommand {\TextLifecycleNonArgumentsd} {\code{AA}\xspace} 
\newcommand {\PercLifecycleNonArgumentsd} {2\%\xspace} 
\newcommand {\TextLifecycleNonArgumentse} {\code{F}\xspace} 
\newcommand {\PercLifecycleNonArgumentse} {2\%\xspace} 
\newcommand {\TextLifecycleNonArgumentsf} {\code{S}\xspace} 
\newcommand {\PercLifecycleNonArgumentsf} {0.2\%\xspace} 
\newcommand {\CountConSenArgForce} {651.7\,K\xspace} 
\newcommand {\PercConSenArgForceSuccess} {86.3\%\xspace} 
\newcommand {\PercConSenArgForceFailure} {13.7\%\xspace} 
\newcommand {\CodeConSenSymbola} {\code{plot}\xspace} 
\newcommand {\CountConSenSymbola} {971\xspace} 
\newcommand {\CodeConSenSymbolb} {\code{log}\xspace} 
\newcommand {\CountConSenSymbolb} {821\xspace} 
\newcommand {\CodeConSenSymbolc} {\code{c}\xspace} 
\newcommand {\CountConSenSymbolc} {461\xspace} 
\newcommand {\CodeConSenSymbold} {\code{legend}\xspace} 
\newcommand {\CountConSenSymbold} {334\xspace} 
\newcommand {\CodeConSenSymbole} {\code{file}\xspace} 
\newcommand {\CountConSenSymbole} {221\xspace} 
\newcommand {\CodeConSenSymbolf} {\code{length}\xspace} 
\newcommand {\CountConSenSymbolf} {209\xspace} 
\newcommand {\CodeConSenSymbolg} {\code{formula}\xspace} 
\newcommand {\CountConSenSymbolg} {204\xspace} 
\newcommand {\CodeConSenSymbolh} {\code{scale}\xspace} 
\newcommand {\CountConSenSymbolh} {188\xspace} 
\newcommand {\CodeConSenSymboli} {\code{t}\xspace} 
\newcommand {\CountConSenSymboli} {184\xspace} 
\newcommand {\CodeConSenSymbolj} {\code{names}\xspace} 
\newcommand {\CountConSenSymbolj} {182\xspace} 
\newcommand {\CodeConSenSymbolk} {\code{round}\xspace} 
\newcommand {\CountConSenSymbolk} {130\xspace} 
\newcommand {\CodeConSenSymboll} {\code{title}\xspace} 
\newcommand {\CountConSenSymboll} {129\xspace} 
\newcommand {\CodeConSenSymbolm} {\code{order}\xspace} 
\newcommand {\CountConSenSymbolm} {83\xspace} 
\newcommand {\CodeConSenSymboln} {\code{drop}\xspace} 
\newcommand {\CountConSenSymboln} {82\xspace} 
\newcommand {\CodeConSenSymbolo} {\code{which}\xspace} 
\newcommand {\CountConSenSymbolo} {82\xspace} 
\newcommand {\CodeConSenSymbolp} {\code{grid}\xspace} 
\newcommand {\CountConSenSymbolp} {76\xspace} 
\newcommand {\CodeConSenSymbolq} {\code{class}\xspace} 
\newcommand {\CountConSenSymbolq} {74\xspace} 
\newcommand {\CodeConSenSymbolr} {\code{print}\xspace} 
\newcommand {\CountConSenSymbolr} {74\xspace} 
\newcommand {\CodeConSenSymbols} {\code{ncol}\xspace} 
\newcommand {\CountConSenSymbols} {71\xspace} 
\newcommand {\CodeConSenSymbolt} {\code{dim}\xspace} 
\newcommand {\CountConSenSymbolt} {69\xspace} 
\newcommand {\CodeConSenSymbolu} {\code{max}\xspace} 
\newcommand {\CountConSenSymbolu} {67\xspace} 
\newcommand {\CodeConSenSymbolv} {\code{format}\xspace} 
\newcommand {\CountConSenSymbolv} {66\xspace} 
\newcommand {\CodeConSenSymbolw} {\code{sort}\xspace} 
\newcommand {\CountConSenSymbolw} {60\xspace} 
\newcommand {\CodeConSenSymbolx} {\code{nrow}\xspace} 
\newcommand {\CountConSenSymbolx} {57\xspace} 
\newcommand {\CodeConSenSymboly} {\code{list}\xspace} 
\newcommand {\CountConSenSymboly} {53\xspace} 
\newcommand {\CodeConSenSymbolz} {\code{rug}\xspace} 
\newcommand {\CountConSenSymbolz} {49\xspace} 
\newcommand {\CodeConSenSymbolA} {\code{matrix}\xspace} 
\newcommand {\CountConSenSymbolA} {48\xspace} 
\newcommand {\CodeConSenSymbolB} {\code{clean}\xspace} 
\newcommand {\CountConSenSymbolB} {47\xspace} 
\newcommand {\CodeConSenSymbolC} {\code{start}\xspace} 
\newcommand {\CountConSenSymbolC} {45\xspace} 
\newcommand {\CodeConSenSymbolD} {\code{unique}\xspace} 
\newcommand {\CountConSenSymbolD} {43\xspace} 
\newcommand {\PercSingleGCPromises} {99.5\%\xspace} 
\newcommand {\PercMultipleGcPromises} {0.5\%\xspace} 
\newcommand {\PercMultipleGcEscapedArgumentPromises} {0.08\%\xspace} 
\newcommand {\PercMultipleGcUnescapedArgumentPromises} {23.2\%\xspace} 
\newcommand {\PercMultipleGcNonArgumentPromises} {76.7\%\xspace} 
\newcommand {\CountEffectPromises} {16.5\,M\xspace} 
\newcommand {\CountEffectSamePromises} {6.8\,M\xspace} 
\newcommand {\PercEffectSamePromises} {41.2\%\xspace} 
\newcommand {\CountEffectOtherPromises} {9.6\,M\xspace} 
\newcommand {\PercEffectOtherPromises} {58.3\%\xspace} 
\newcommand {\CountEffectLexicalPromises} {80.2\,K\xspace} 
\newcommand {\PercEffectLexicalPromises} {0.5\%\xspace} 
\newcommand {\PercEffectDirectSamePromises} {98\%\xspace} 
\newcommand {\PercEffectDirectOtherPromises} {0.1\%\xspace} 
\newcommand {\PercEffectDirectLexicalPromises} {0.08\%\xspace} 
\newcommand {\PercPromiseZeroDepth} {79.7\%\xspace} 
\newcommand {\PercPromiseOneDepth} {17.1\%\xspace} 
\newcommand {\PercPromiseMultipleDepth} {3.2\%\xspace} 
\newcommand {\PercPromiseFunctionCallCallerEvaluation} {7.0\%\xspace} 
\newcommand {\PercPromiseFunctionCallRemoteEvaluation} {12.3\%\xspace} 
\newcommand {\PercPromiseFunctionCallNonEvaluation} {80.7\%\xspace} 
\newcommand {\CountMutualForcing} {336.9\,M\xspace} 
\newcommand {\PercMutualForcing} {0.2\%\xspace} 
\newcommand {\CountMutualForcingFunctions} {3.9\,K\xspace} 
\newcommand {\CountMutualForcingArgumentPositions} {4.8\,K\xspace} 
\newcommand {\CountNonLocalFunctions} {16\xspace} 
\newcommand {\CountNonLocalArguments} {297.4\,M\xspace} 
\newcommand {\CountSThreePromiseForce} {671.5\,M\xspace} 
\newcommand {\CountSFourPromiseForce} {536.2\,M\xspace} 

\usepackage{my_style}

\definecolor{LightGray}{rgb}{.92,.92,.92}
\definecolor{Gray}{rgb}{.3,.3,.3}
\definecolor{DarkGray}{rgb}{.5,.5,.5}

\lstdefinestyle{R}{ %
  language=R,
  deletekeywords={new, env, equal, c, runif, trace, args, exp, t, all, get,
    names, is, environment, class, substitute, expression, list, null, Internal,
    sample, diag, length, rep, nrow, stop, offset, pmax, Machine,
    double, parent, frame, par, methods, end, dir, apply, deparse, missing,
    plot, as, integer, character, inherits, numeric, paste, eval, quote, call,
    formula, df, log, sum, c, local, legend, file, scale, round, title, order,
    drop, which, grid, print, ncol, dim, max, format, sort, rug, matrix, start,
    unique, mean, df, attr, do, power},
  otherkeywords={},
  breaklines=true
}

\newcommand{\code}[1]{\lstinline[style=R]|#1|\xspace}
\renewcommand{\c}[1]{\lstinline[style=R]|#1|\xspace}

\setcopyright{rightsretained}
\acmPrice{}
\acmDOI{10.1145/3360579}
\acmYear{2019}
\copyrightyear{2019}
\acmJournal{PACMPL}
\acmVolume{3}
\acmNumber{OOPSLA}
\acmArticle{153}
\acmMonth{10}
\begin{document}
\title{On the Design, Implementation, and Use of Laziness in R}

\author{Aviral Goel}\affiliation{\institution{Northeastern University}\country{USA}}
\author{Jan Vitek}\affiliation{\institution{Czech Technical University and Northeastern University}\country{USA}}
\authorsaddresses{} 
\renewcommand{\shortauthors}{Goel, Vitek}

\begin{abstract} 
The R programming language has been lazy for over twenty-five years. This
paper presents a review of the design and implementation of call-by-need in
R, and a data-driven study of how generations of programmers have put
laziness to use in their code. We analyze \CountCorpusPackages packages and
observe the creation of \CountProm promises. Our data suggests that there is
little supporting evidence to assert that programmers use laziness to avoid
unnecessary computation or to operate over infinite data structures. For the
most part R code appears to have been written without reliance on, and in
many cases even knowledge of, delayed argument evaluation. The only
significant exception is a small number of packages which leverage call-by-need
for meta-programming.
\end{abstract}



\maketitle

\section{Introduction}

Since its inception, in 1993, R has had a call-by-need semantics.  When a
function is invoked its arguments are packaged up into \emph{promises} which
are evaluated on demand. The values obtained by evaluating those promises
are memoized to avoid the need for recomputation. Thus the following
definition when called with \code{a+b} and \code{d+d}

\vspace{-1mm}
\begin{lstlisting}[style=R]
  f <- function(x,y) x + x
\end{lstlisting}

\noindent
evaluates \code{a+b} and does not evaluate \code{d+d}. With an estimated two
million users world-wide~\cite{eco11}, R is the most widely used lazy
functional programming language in existence. The R community has developed
an extensive body of reusable, documented, tested and maintained code. The
CRAN open source repository, the main source for R packages, hosts over
16,000 packages with an average of 6 new ones added every
day~\cite{LIgges2017}.  Bioconductor~\cite{bioc}, another open source R
package repository, hosts over 1,700 packages for genomic data analysis.

It is fascinating to observe that R's laziness mostly remains secret. The
majority of end-users are unaware of the semantics of the language they
write code in. Anecdotally, this holds even for colleagues in the
programming language community who use R casually. Moreover, we do not know
of any studies of the design and efficacy of call-by-need in R. With twenty-five
years of practical experience with laziness, some lessons can surely be drawn.

\citet{Hudak89} defined lazy evaluation as the implementation of
normal-order reduction in which recomputation is avoided. He went on to
enumerate two key benefits for programmers: (1) Sub-computations are only
performed if they are needed for the final result; (2) Unbounded data
structures include elements which are never materialized.  Haskell is a
language designed and implemented to support lazy evaluation, its compiler
has optimization passes that remove some of the overhead of delayed
evaluation, and its type system allows laziness to co-exist with side
effects in an orderly manner.

R differs from Haskell in its approach to lazy evaluation.  The differences
are due in part to the nature of the language and in part to the goals of
its designers. As R frequently calls into legacy C and Fortran libraries,
performance dictates that the memory layout of R objects be consistent with
the expectations of those libraries. For statistical and mathematical codes,
this mostly means array of primitive types, integer and floating point,
should be laid out contiguously using machine representations for primitives.
Interoperability is thus the reason for builtin datatypes being strict, and
consequently for mostly giving up on the second benefit of laziness right
out of the gate. As for the first benefit, it goes unfulfilled because R
tries to only be as lazy as it needs. In numerous places, design choices
limit its laziness in favor of a more predictable order of execution; this
is compounded by a defensive programming style adopted in many packages
where arguments are evaluated to obtain errors early.

Given the above, one may wonder \emph{why bother being lazy?} In particular,
when this implies run-time costs that can be significant as R does not
optimize laziness. Communications with the creators of R suggests
call-by-need was added to support meta-programming and, in particular,
user-defined control structures. While R was inspired by Scheme, the
latter's macro-based approach to meta-programming was not adopted. Macros
were deemed too complex for users. Instead, R offers a combination of
call-by-need and reflection. Call-by-need postpones evaluation so that
reflection can inspect and modify expressions, either changing their binding
environment or their code.  This approach was deemed sufficiently expressive
for the envisioned use-cases and, unlike macros, it was not limited to
compile-time redefinition---an important consideration for an interactive
environment. Evaluating how that choice worked out is an interesting but
separate question.

This paper describes the design and implementation of call-by-need in the
production R virtual machine, and provides data about its use in
practice. The latter is obtained by dynamic analysis of a corpus of programs
hosted in the CRAN and Bioconductor software repositories. Dynamic analysis
is limited to behaviors that it observes, low code coverage may affect the
results. Values that appear not to be evaluated (or always evaluated in the
same order) may be evaluated (or evaluated in different orders) in control
flow branches that were not exercised.  We are encouraged by the fact that
\citet{issta18} reports line-level code coverage of over 60\% for a similar
corpus. To mitigate this threat to validity, we perform a qualitative
analysis of a sample of our corpus.  We claim the following contributions:
\begin{itemize}
\item We describe the design and implementation of call-by-need in R and
  give a small-step operational semantics for a subset of the language that
  includes promises, \code{eval}, \code{substitute} and
  \code{delayedAssign}.
\item We release an open source, carefully optimized, dynamic analysis
  pipeline consisting of an instrumented R interpreter and data analysis
  scripts. 
\item We present results about the strictness of functions, their possible
  evaluation orders, as well as the life cycle of promises from the analysis of
  \ProgramCount scripts exercising code from \CountCorpusPackages packages.
\end{itemize}

\noindent
Our semantics has not been validated by testing.
Our results were obtained with version \GNURVERSION of \GNUR and packages
retrieved on August 1st, 2019 from CRAN and Bioconductor.  Our software and
data was validated as Functional and Reusable and is available in open
source from:

\vspace{2mm}
\hspace{3cm}\url{https://doi.org/10.5281/zenodo.3369573}
\vspace{2mm}

\renewcommand{\k}[1]{{\tt #1}\xspace}
\newpage

\section{Background on R}\label{sec:rlang}

The R project is a tool for implementing sophisticated data analysis
algorithms. At heart, R is a \emph{vectorized, dynamic, lazy, functional and
  object-oriented} programming language with a rather unusual combination of
features~\cite{ecoop12}, designed to be easy to learn by non-programmers and
to enable rapid development of new statistical methods.  The language was
created in 1993 by~\citet{R96} as a successor to an earlier language for
statistics named S~\cite{S88}. We focus on features relevant to this work.

In R, most data types are vectorized. Values are constructed by the
\code{c(...)} function: \code{c("hi","ho")} creates a vector of two
strings. The language does not differentiate scalars from vectors, thus
\code{1==c(1)}. In order to enable equational reasoning, R copies most values
accessible through multiple variables if they are written to.  Consider the
\code{swap} function which exchanges two elements in a vector and returns
the modified vector:

\begin{lstlisting}[style=R]
> swap <- function(x, i, j) { t<-x[i]; x[i]<-x[j]; x[j]<-t; x }
> v <- c(1,2,3)
> swap(v,1,3)
\end{lstlisting}\vspace{1mm}
\vspace{2mm}

\noindent
The argument vector \k v is shared, as it is aliased by \k x in the
function. Thus, when \code{swap} first writes to \k x at offset \k i, the
vector is copied, leaving \k v unchanged. It is the copy that is returned by
\code{swap}. Behind the scenes, a reference count is maintained for
all objects. Aliasing a value increases the count. Any update of a value
with a count larger than one triggers a copy. One motivation for this design
was to allow users to write iterative code that updates vectors in place.  A
loop that updates all elements of an array will copy at most once.

Every linguistic construct is desugared to a function call, even control
flow statements, assignments, and bracketing. Furthermore, all functions can
be redefined in packages or user code. This makes R both flexible and
challenging to compile~\cite{rmeltbrains}. A function definition can include
default expressions for parameters, these can refer to other parameters. R
functions are higher-order. The following snippet declares a function \k f
which takes a variable number of arguments (triple dots are a vararg), whose
parameters \k x and \k y, if missing, evaluate to \k y and \code{3*x}. The
function returns a closure.

\begin{lstlisting}[style=R]
> f <- function(x=y, ..., y=3*x) { function(z) x+y+z }
\end{lstlisting}\vspace{1mm}

\noindent
Function \k f can be called with no arguments \code{f()}, with a single
argument \code{f(3)}, with named arguments \code{f(y=4,x=2)} and with a
variable number of arguments, \code{f(1,2,3,4,y=5)}.

Values can be tagged by user-defined attributes. For instance, one can
attach the attribute \k{dim} to the value \code{x<-c(1,2,3,4)} by
evaluating \code{attr(x,"dim")<-c(2,2)}.  Once done, arithmetic functions
will treat \k x as a \k{2x2} matrix. Another attribute is \k{class} which can
be bound to a list of names. For instance, \code{class(x)<-"human"}, sets
the class of \k{x} to \k{human}.  Attributes are used for object-oriented
dispatch. The ``S3 object system'' supports single dispatch on the class of
the first argument of a function, whereas the ``S4 object system'' allows
dispatch on all arguments. These names refer to the version of the S
language which introduced them.  Popular data types, such as data frames,
also leverage attributes. A data frame is a list of vectors with
\code{class} and \code{colname} attributes.

R supports reflection and meta-programming. The \code{substitute(e,envir)}
function yields the parse tree of the expression \code{e} after performing
substitutions defined by the bindings in \code{envir}.
\vspace{-1mm}
\begin{lstlisting}[style=R]
> substitute(expression(a + b), list(a = 1)))  
expression(1 + b)
\end{lstlisting}
\vspace{2mm}

\noindent
R allows programmatic manipulation of parse trees, which are themselves
first class objects. They are evaluated using the \rcode{eval(e,envir)}
function. Environment on the call stack can also be accessed.

\newpage
\section{Call-by-need in R}

The combination of side effects, frequent interaction with C, and absence of
types has pushed R to be more eager than other lazy languages.  This section
reviews the design and implementation of laziness.  In R, arguments to a
user-defined function are bundled into a thunk called a
\emph{promise}. Logically, a promise combines an expression's code, its
environment, and its value.  To access the value of a promise, one must
\emph{force} it. Forcing a promise triggers evaluation and the
computed value is captured for future reference. The following snippet defines a function
\rcode{f} that takes argument \k x and returns \code{x+x}. When called with
an argument that has the side effect of printing to the console, the side
effect is performed once as the second access to the promise is cached.
\begin{lstlisting}[style=R]
> f <- function(x) x+x
> f( {print("Hi!");2} )
"Hi!"
4
\end{lstlisting}
\vspace{2mm}

\noindent
Promises associated to parameters' default values have access to all
variables in scope, including other parameters.  Promises cannot be forced
recursively, that is an error.  Promises are mostly encapsulated and hidden
from user code. R only provides a small interface for operating on promises:
\begin{itemize}
\item {\bf\small\code{delayedAssign(x,exp,eenv,aenv)}}: create a promise
  with body \code{exp} and binds it to variable \k x (where \k x is a
  symbol). Environment \code{eenv} is used to evaluate the promise, and
  \code{aenv} is used to perform the assignment.
\item {\bf\small\code{substitute(e,env)}}: substitutes variables in \k e
  with their values found in environment \code{env}, returns an
  \code{expression} (a parse tree).
\item {\bf\small\code{force(x)}}: forces the promise \k x. This replaces a
  common programming idiom, \code{x<-x}, which forces \k x by assigning it
  to itself.
\item {\bf\small\code{forceAndCall(n,f,...)}}: call \code{f} with the
  arguments specified in the varargs of which the first \code{n} are forced
  before the call.
\end{itemize}

\noindent
While R does not provide built-in lazy data structures, they can be encoded.
Figure~\ref{ll} shows a lazy list that uses environments as structs;
environments have reference semantics. R provides syntactic sugar for
looking up variables (\$), functions for creating environment out of lists
(\code{list2env}) and for capturing the current environment
(\k{environment}). The singleton \k{nil} has a \k{tag} that is tested in the
\k{empty} function.  A new list is created by \k{cons}; it returns its
environment in which \k h and \k t are bound to promises. The \k{head} and
\k{tail} functions retrieve the contents of \k h and \k t respectively.
This example also illustrates how promises can be returned from their
creation environment, namely by being protected by an environment.

\begin{figure}[!h]
\begin{lstlisting}[style=R]
  nil   <- list2env(list(tag="nil"))
  empty <- function(l) l$tag=="nil"
  cons  <- function(h,t) environment()
  head  <- function(l) l$h
  tail  <- function(l) l$t
\end{lstlisting} 
\caption{Lazy list in R}\label{ll}
\end{figure}

\noindent
R evaluates promises aggressively. The sequencing operator \k{a;b} will
evaluate both \k a and \k b, assignment \code{x<-a} evaluates \k a, and
\code{return} also triggers evaluation.  In addition, many core functions
are strict.  R has two kinds of functions that are treated specially:

\begin{itemize}
\item \textit{Builtins}: There are \BUILTINCOUNT builtins, typically written
  in C, providing efficient implementations of numerical methods and other
  mathematical functions. The argument lists of builtins are evaluated
  eagerly.
\item \textit{Specials}: There are \SPECIALCOUNT specials used to implement
  core language features such as loops, conditionals, bracketing, etc. These
  functions take expressions (parse trees) which are evaluated in the
  calling environment or in a specially constructed environment.
\end{itemize}

\noindent
Builtins and specials are exposed as functions to the surface language
either directly or through wrappers which perform pre-processing of arguments
before passing them to these functions.

We would be remiss if we did not mention context-sensitive lookup, one of
the unusual features of R. When looking up a variable \k x, in head position
of a function call, e.g., \code{x(...)}, R finds the first definition of \k
x in a lexically enclosing environment, if \k x is bound to a closure, that
closure is returned. Otherwise, lookup continues in the enclosing scope. A
corollary of function lookup is that it forces promises encountered along the
way.

\subsection{Implementing Promises}

A promise has four slots: \k{exp}, \k{env}, \k{val} and \k{forced}. The
\k{exp} slot contains a reference to the code of the promise, the \k{env}
refers to environment in which the promise was originally created. The
\k{val} slot holds the result of evaluating the \k{exp}.  The \k{forced}
flag is used to avoid recursion.  When a promise is accessed, the \k{val}
slot is inspected first. If it is not empty, that value is
returned. Otherwise, \k{forced} is checked, and if it is set an exception is
thrown. The \k{forced} flag is updated and the expression is evaluated in the
specified environment.  Once the evaluation finishes, the \k{val} slot is
bound to the result, the \k{env} slot is cleared to allow the environment to
be reclaimed, and the \k{forced} flag is unset.  The implementation does
little to optimize promises. In some cases, a promise can be created
pre-forced with a value pre-assigned. The \GNUR implementation recently
added a bytecode compiler, this compiler eliminates promises when they
contain a literal~\cite{tie19}.

\medskip

\medskip

\section{Semantics}

This section describes a small-step operational semantics in the style
of~\cite{Wright92} for a core R language with promises. Our goal is to
provide an easy to follow---the entire semantics fits on a page---but
precise---unlike the above prose description---account of R's call-by-need
semantics.  We build upon the semantics of Core R~\cite{ecoop12}, but omit
vectors and out-of-scope assignments. Instead, we add delayed assignment,
default values for arguments, \code{substitute} and \code{eval}. To support
these features we add strings as a base type and the ability to capture the
current environment.

\newcommand{\CC}{\ensuremath{\mathbb C}\xspace}
\begin{figure}[!b]\footnotesize\hrulefill\vspace{1mm}

\begin{tabular}{L@{}L@{~}L}
\EXP~\ ~ ::=~\ ~ \STR  \OR \VAR
      \OR \concat\EXP\EXP 
      \OR \assign\VAR\EXP
      \OR \function\VAR\EXP\EXP 
      \OR \VAR(\EXP)
      \OR \VAR()
      \OR \env
      \OR \subst\VAR
      \OR \eval\EXP\EXP
      \OR \delay\VAR\EXP\EXP
\end{tabular}

~\\

{\small$
\CC ~\ ~::=~\ ~ [] \OR 
      \OR \concat\CC\EXP
      \OR \concat v\CC
      \OR \assign\VAR\CC
      \OR \CC(\EXP)
      \OR \CC()
      \OR \eval\CC\EXP
      \OR \eval v\CC
      \OR \delay\VAR\EXP\CC $}

~\\

\begin{tabular}{L@{~\ ~}cLll}
 \FRAME &::=& \EMPTYENV \OR \extend{\FRAME}{\VAR}{\REF} & \it frames&
~\ ~\ ~\ ~\ ~\ ~\ ~\ ~\ ~\ ~\ ~\ ~\ ~\ ~\ ~\ ~\ ~\ ~\ ~\ ~\ ~\ ~\ ~\ ~\ ~\ ~\ ~\ ~\ ~\ ~\ ~\ ~\ ~\ ~\ ~\ ~\ ~\ ~\ ~\ ~\ ~\ ~\ ~\ ~\ ~\ ~\ ~\ ~\ ~\ ~\ 
\\
 \ENV &::=& \EMPTYENV \OR \seqtwo\REF\ENV & \it environments \\
 \STACK &::=& \EMPTYSTACK \OR \seqtwo{\EXP\,\ENV}\STACK&\it stacks \\
\HEAP &::=& \EMPTYHEAP \OR \extend\HEAP\REF\VAL 
   \OR \extend\HEAP\REF\FRAME
   \OR \extend\HEAP\REF{\quadruplet\VAL\EXP\ENV\FLAG} & \it heap \\
 \FLAG &\enspace::=\enspace& \SEEN \OR \UNSEEN  &\it forced flag
\end{tabular}

~\\[-3mm]
\hrulefill

\begin{mathpar}
\inferrule* [Lab=\tiny\irlabel{Fun}] {
   v = \closure\VAR\EXP{\sglprime\EXP}\ENV
} {
  \dblrel{\seqtwo{\C{\function\VAR\EXP{\sglprime\EXP}}\,\ENV}\STACK}\HEAP
         {\seqtwo{\C v\,\ENV}\STACK}\HEAPp
} \and
\inferrule* [Lab=\tiny\irlabel{Concat}] {
 } {
  \dblrel{\seqtwo{\C{\concat{\STR}{\STR'}}\, \ENV}\STACK}\HEAP
         {\seqtwo{\C{\STR\STR'}\,\ENV}\STACK}\HEAP
} \and
\inferrule* [Lab=\tiny\irlabel{Assign}] {
  \ENV = \seqtwo\REFp\ENVp \\
  \HEAP(\REFp) = \FRAME \\\\
  \FRAMEp = \extend\FRAME\VAR v \\
  \HEAPp = \extend\HEAP\REFp\FRAMEp
} {
  \dblrel {\seqtwo{\C{\assign\VAR v}\,\ENV}\STACK}\HEAP
          {\seqtwo{\C{v}\,\ENV}\STACK}\HEAPp
} \and
\inferrule* [Lab=\tiny\irlabel{Delay}] {
  \HEAP(\REF) =  \seqtwo\REFp\ENVp \\
  \HEAP(\REFp) = \FRAME \\
  \fresh\REFpp \\\\
  \FRAMEp = \extend\FRAME\VAR{\promval{\REFpp}} ~\ ~\ ~\ 
  \HEAPp = \extend
            {\extend\HEAP\REFpp{\quadruplet\NULLREF\EXP\ENV\UNSEEN}}\REFp\FRAMEp
} {
  \dblrel {\seqtwo{\C{\delay\VAR\EXP{\envval{\REF}}}\,\ENV}\STACK}\HEAP
          {\seqtwo{\C{\envval{\REF}}\,\ENV}\STACK}\HEAPp
} \and
    \inferrule* [Lab=\tiny\irlabel{Env}] {
      \fresh\REF \\
      \HEAPp = \extend\HEAP\REF\ENV
    } {
      \dblrel {\seqtwo{\C\env\,\ENV}\STACK}\HEAP
              {\seqtwo{\C{\envval\REF}\,\ENV}\STACK}\HEAPp
    } \and
\inferrule* [Lab=\tiny\irlabel{Subst}] {
  \get\HEAP\ENV\VAR = v\\
  v = \quadruplet\_\EXP\_\_ \\
  \str\EXP = \STR
} {
  \dblrel{\seqtwo{\C{\subst\VAR}\,\ENV}\STACK}\HEAP
         {\seqtwo{\C\STR\,\ENV}\STACK}\HEAP
} \and
\inferrule* [Lab=\tiny\irlabel{Eval}] {
  \EXP = \eval\STR{\envval\REF}\\
  \mathit{parse}(\STR) = \EXPp \\
  \HEAP(\REF) = \ENVp
} {
  \dblrel{\seqtwo{\C\EXP\,\ENV}\STACK}\HEAP
         {\seqthree{\EXPp\,\ENVp}{\C{\EXP}\,\ENV}\STACK}\HEAP
} \and
\inferrule* [Lab=\tiny\irlabel{EvalRet}] {
  \EXP = \eval\STR{\envval\REFp}\\
} {
  \dblrel{\seqthree{v\,\ENVp}{\C\EXP\,\ENV}\STACK}\HEAP
         {\seqtwo{\C v\,\ENV}\STACK}\HEAP
} \and
\end{mathpar}

\begin{mathpar}
\inferrule* [Lab=\tiny\irlabel{Invk1}] {
  v =\closure\VAR\EXPp\EXPpp\ENVp\\
  \fresh{\REF,\REFp} \\\\
  \ENVpp = \seqtwo\REF\ENVp \\
  \HEAPp = \extend\HEAP\REF\FRAME\\\\
  \FRAME = \extend{}\VAR\REFp \\
  \HEAPpp = \extend\HEAPp\REFp{\quadruplet\NULLREF\EXP\ENV\UNSEEN}
} {
  \dblrel{\seqtwo{\C{v(\EXP)}\,\ENV}\STACK}\HEAP
         {\seqthree{\EXPpp\,\ENVpp}{\C{v(\EXP)}\,\ENV}\STACK}\HEAPpp
} \and
\inferrule* [Lab=\tiny\irlabel{Invk0}] {
  v=\closure\VAR\EXPp\EXPpp\ENVp\\
  \fresh{\REF,\REFp} \\\\
  \ENVpp = \seqtwo\REF\ENVp \\
  \HEAPp = \extend\HEAP\REF\FRAME\\\\
   \FRAME = \extend{}\VAR\REFp \\
   \HEAPpp = \extend\HEAPp\REFp{\quadruplet\NULLREF\EXPp\ENVpp\UNSEEN}
} {
  \dblrel{\seqtwo{\C{v()}\,\ENV}\STACK}\HEAP
         {\seqthree{\EXPpp\,\ENVpp}{\C{v()}\,\ENV}\STACK}\HEAPpp
} \and
    \inferrule* [Lab=\tiny\irlabel{Ret1}] {
    } {
      \dblrel {\seqthree{v\,\ENVp}{\C{v'(\EXP)}\,\ENV}\STACK}\HEAP
              {\seqtwo{\C{v}\,\ENV}\STACK}\HEAP
    }\and
    \inferrule* [Lab=\tiny\irlabel{Ret0}] {
    } {
      \dblrel {\seqthree{v\,\ENVp}{\C{v'()}\,\ENV}\STACK}\HEAP
              {\seqtwo{\C{v}\,\ENV}\STACK}\HEAP
    }\and
\inferrule* [Lab=\tiny\tiny\irlabel{Lookup}] {
  \get\HEAP\ENV\VAR = v \\
  v \not= \promval\REF
} {
  \dblrel {\seqtwo{\C\VAR\,\ENV}\STACK}\HEAP
          {\seqtwo{\C v \,\ENV}\STACK}\HEAP
} \and
\inferrule* [Lab=\tiny\tiny\irlabel{Lookup2}] {
  \get\HEAP\ENV\VAR = \promval\REF
} {
  \dblrel {\seqtwo{\C\VAR\,\ENV}\STACK}\HEAP
          {\seqthree{\promval\REF\,\ENV}{\C\VAR\,\ENV}\STACK}\HEAP
} \and
\inferrule* [Lab=\tiny\irlabel{Force}] {
  \HEAP(\REF) =\quadruplet\NULLREF\EXP\ENVp\UNSEEN \\
  \HEAPp = \extend\HEAP\REF{\quadruplet\NULLREF\EXP\ENVp\SEEN}
} {
  \dblrel{\seqtwo{\promval\REF\,\ENV}\STACK}\HEAP
         {\seqthree{\EXP\,\ENVp}{\promval\REF\,\ENV}\STACK}\HEAPp
} \and
\inferrule* [Lab=\tiny\irlabel{ReadVal}] {
  \HEAP(\REF)=\quadruplet v\EXP\EMPTYENV\UNSEEN
} {
  \dblrel {\seqtwo{\promval\REF\,\ENV}\STACK}\HEAP
          {\seqtwo{v\,\ENV}\STACK}\HEAP
} \and\
\inferrule* [Lab=\tiny\irlabel{Memo}] {
  v \not=\promval\REFpp ~\ ~\ ~\  
  \HEAP(\REF) = \quadruplet\NULLREF\EXP\ENVp\SEEN ~\ ~\ ~\ ~~
  \HEAPp = \extend\HEAP\REF{\quadruplet v\EXP\EMPTYENV\UNSEEN}
} {
   \dblrel {\seqthree{v\,\ENVp}{\promval\REF\,\ENV}\STACK}\HEAP
           {\seqtwo{v\,\ENV}\STACK}\HEAPp
} \and
\inferrule* [Lab=\tiny\irlabel{RetProm}] {
  v \not=\promval\REF
} {
   \dblrel {\seqthree{v\,\ENVp}{\C\VAR\,\ENV}\STACK}\HEAP
           {\seqtwo{\C v\,\ENV}\STACK}\HEAPp
} \and
\end{mathpar}
\\[-2mm]
\hrulefill
\vspace{2mm}
\begin{mathpar}
    \inferrule* {
      \ENV = \seqtwo\REF\ENVp  \hspace{3mm}
      \HEAP(\REF) = \FRAME \hspace{3mm}
      \FRAME(\VAR) = v 
    } {
      \get\HEAP\ENV\VAR = v
    } \and
    \inferrule* {
      \ENV = \seqtwo\REF\ENVp  \hspace{3mm}
      \HEAP(\REF) = \FRAME  \hspace{3mm}
      \VAR \notin \domain\FRAME
    } {
      \get\HEAP\ENV\VAR = \get\HEAP\ENVp\VAR
    } \and
\end{mathpar}
\hrulefill
\caption{Syntax and Semantics}\label{sem}\end{figure}

Figure~\ref{sem} gives the syntax and semantics of our calculus.  The
surface syntax includes terms for strings, variables, string concatenation,
assignment, function declaration, function invocation (one and zero argument
functions), environment capture, substitution, eval, and delayed assignment.
The syntax is extended with additional terms used during reduction where
variables and expressions can be replaced by values ranged over by
meta-variable $v$ which can be one of a string (\STR), a closure
\closure\VAR\EXP\EXP\ENV, an environment (\envval\REF), or a promise
(\promval\REF).  Mutable values are heap allocated and ranged over by
meta-variable \REF. We use the \NULLREF value to denote an invalid
reference.

The reduction relation is of the form $\STACK\;\ENV \rightarrow
\STACK';\ENVp$ where the stack \STACK is a collection of
expression-environment pairs (\EXP\,\ENV) and the heap maps variables to
values. Frames are mutable and can be shared between closures, so they are
stack allocated.  Promises are quadruples $\quadruplet\VAL\EXP\ENV\FLAG$
where $v$ is the cached result of evaluating the body \EXP in environment
\ENV. \FLAG is a status flag where \UNSEEN indicates we have started
evaluating the promise.  Evaluation contexts \CC are deterministic.
Following R, some builtin operations such as string concatenation are
strict.  But function calls are lazy in their argument, the expression in
\VAR(\EXP) remains untouched by the context.  We omit the definition of the
functions \emph{parse} and \emph{string} which, respectively, turn strings
into expressions and vice-versa.  The expression \emph{fresh} is used to
obtain new heap references.

The semantics is given by the following rules.  Rules \irlabel{Fun} and
\irlabel{Concat} deal with creating a closure in the current environment and
concatenating string values.  Rule \irlabel{Assign} will add a mapping from
variable \VAR to value $v$ in the current frame. As frames are allocated in
the heap, this updates the heap.  Rule \irlabel{Delay} performs a delayed
assignment, that is to say, an assignment that does not force the right-hand
side. It takes a variable, an expression and an environment in which this
expression will be evaluated.  The rule creates a new unevaluated promise
and binds it to \VAR.  Rule \irlabel{Env} grabs the current environment and
returns it as a value.  Rule \irlabel{Subst} looks up the variable given as
argument, obtains the promise bound to it, extracts its body and deparses it
into a string. Rule \irlabel{Eval} takes a string and an environment, parses
that string into an expression and schedules it for execution. Rule
\irlabel{EvalRet} takes the result of that evaluation and replaces the call
to eval with it.  Rule \irlabel{Invk1} and \irlabel{Invk0} handle
user-defined function calls.  Both rules expect $v$ to be a closure. They
allocate a new promise for \VAR. They differ on the body of
that closure and the environment.  \irlabel{Invk0} has no argument and will
use the default expression \EXP' specified in the function declaration, it
uses the environment where \VAR is bound for evaluation.  Rules
\irlabel{Ret1} and \irlabel{Ret0} are the corresponding returns rules that
replace the call with the computed value.  Rules \irlabel{Lookup} and
\irlabel{Lookup1} are used to read variables from the current
environment. If the result is a promise it is scheduled for execution by
\irlabel{Lookup1}. Rule \irlabel{Force} will actually evaluate a promise
that has been pushed on the stack, if that promise has not yet been
evaluated. It sets the flag to avoid recursive evaluation. Rule
\irlabel{ReadVal} retrieves the value of an already evaluated promise.  Rule
\irlabel{Memo} stores the result of evaluation in the promise and discard
its environment. Finally, rule \irlabel{RetProm} returns from evaluating a
promise by replacing the variable looked up with the result.

\medskip

\subsection{Takeaways} R is rather strict for a lazy language. This manifests
itself in the definition of evaluation contexts. Intuitively, any position
where \CC appears is evaluated strictly in left-to-right order. The key
place where R differs from other lazy languages is that the right-hand side
of assignments is strict. Our semantics does not show data structures, in R
they are all strict. Strictness also shows up in the \irlabel{Ret1} and
\irlabel{Ret0} rules which force the evaluation of the return
expression. Lastly, strictness is enforced in the \irlabel{Lookup2} rule
which does function lookup.  If a promise is returned, it must be evaluated.
Another property that the semantics ensures is that promises are stored in
environments and, whenever they are accessed they are forced. The only way
for a promise to outlive the frame that created it is to be returned as part
of an environment or closure.  Following R, it is possible to create a cycle
in promise evaluation, the expression $(\function\VAR\VAR\VAR)()$ when
evaluated creates a closure and invokes it. The function's body triggers
evaluation of the promise bound to \VAR. Since no argument was provided, the
default expression is evaluated causing a cycle. Like in R, this results in a
stuck state in the semantics.
\newpage
\section{Analysis Infrastructure}

We now explain the infrastructure that assembles the corpus and collects and
analyzes traces.  Our analysis pipeline starts with scripts to download,
extract and install open source R packages.  Next, an instrumented R virtual
machine generates events from program runs. This is followed by an analyzer
that processes the execution traces to generate tabular data files in a
custom binary format. Other scripts post-process the data, compute
statistics, and generate graphs.  The entire pipeline is managed by a Makefile
that invokes an R script to extract runnable code snippets from installed
packages and runs the other steps in parallel. Parallelization is achieved using
GNU Parallel.


\begin{figure}[!h]
\centering
\begin{tikzpicture}
  \definecolor{corpusbg}{HTML}{F5F5F5}
  \definecolor{corpusol}{HTML}{666666}

  \definecolor{tracebg}{HTML}{DAE8FC}
  \definecolor{traceol}{HTML}{6C8EBF}

  \definecolor{reducebg}{HTML}{D5E8D4}
  \definecolor{reduceol}{HTML}{82B366}

  \definecolor{combinebg}{HTML}{FFE6CC}
  \definecolor{combineol}{HTML}{D79B00}

  \definecolor{mergebg}{HTML}{FFF2CC}
  \definecolor{mergeol}{HTML}{D6B656}

  \definecolor{summarizebg}{HTML}{F8CECC}
  \definecolor{summarizeol}{HTML}{B85450}

  \definecolor{reportbg}{HTML}{E1D5E7}
  \definecolor{reportol}{HTML}{9673A6}

  \newcommand{\fstdtpt}[5]{
    \node [inner sep = 0, align = center, below right = 3mm of #5.south west, minimum width = 0.03 \textwidth] (#1) {
      \includegraphics[height=0.02 \textwidth]{#2}
    };
    \node [inner sep = 0, right = 0mm of #1, minimum width = 0.07 \textwidth, text width = 0.07 \textwidth, align = right] (#3) {\scriptsize #4}
  }

  \newcommand{\rstdtpt}[5]{
    \node [inner sep = 0, align = center, below = 3mm of #5, minimum width = 0.03 \textwidth] (#1) {
      \includegraphics[height=0.02 \textwidth]{#2}
    };
    \node [inner sep = 0, right = 0mm of #1, minimum width = 0.07 \textwidth, text width = 0.07 \textwidth, align = right] (#3) {\scriptsize #4}
  }

  \newcommand{\rstdtpth}[5]{
    \node [inner sep = 0, align = center, below = 3mm of #5, minimum width = 0.03 \textwidth] (#1) {
      \includegraphics[width=0.02 \textwidth]{#2}
    };
    \node [inner sep = 0, right = 0mm of #1, minimum width = 0.07 \textwidth, text width = 0.07 \textwidth, align = right] (#3) {\scriptsize #4}
  }

  \tikzstyle{block} = [rectangle, minimum width=.135 \textwidth, minimum height=15pt]

  \newcommand{\fststage}[4]{\node [block, fill = #3] (#1) {\small #2};} 
  \newcommand{\rststage}[5]{\node [block, right = 1mm of #3, fill = #4] (#1) {\small #2};} 

  \fststage{corpus}{Corpus}{corpusbg}{corpusol}
  \fstdtpt{corpustimeicon}{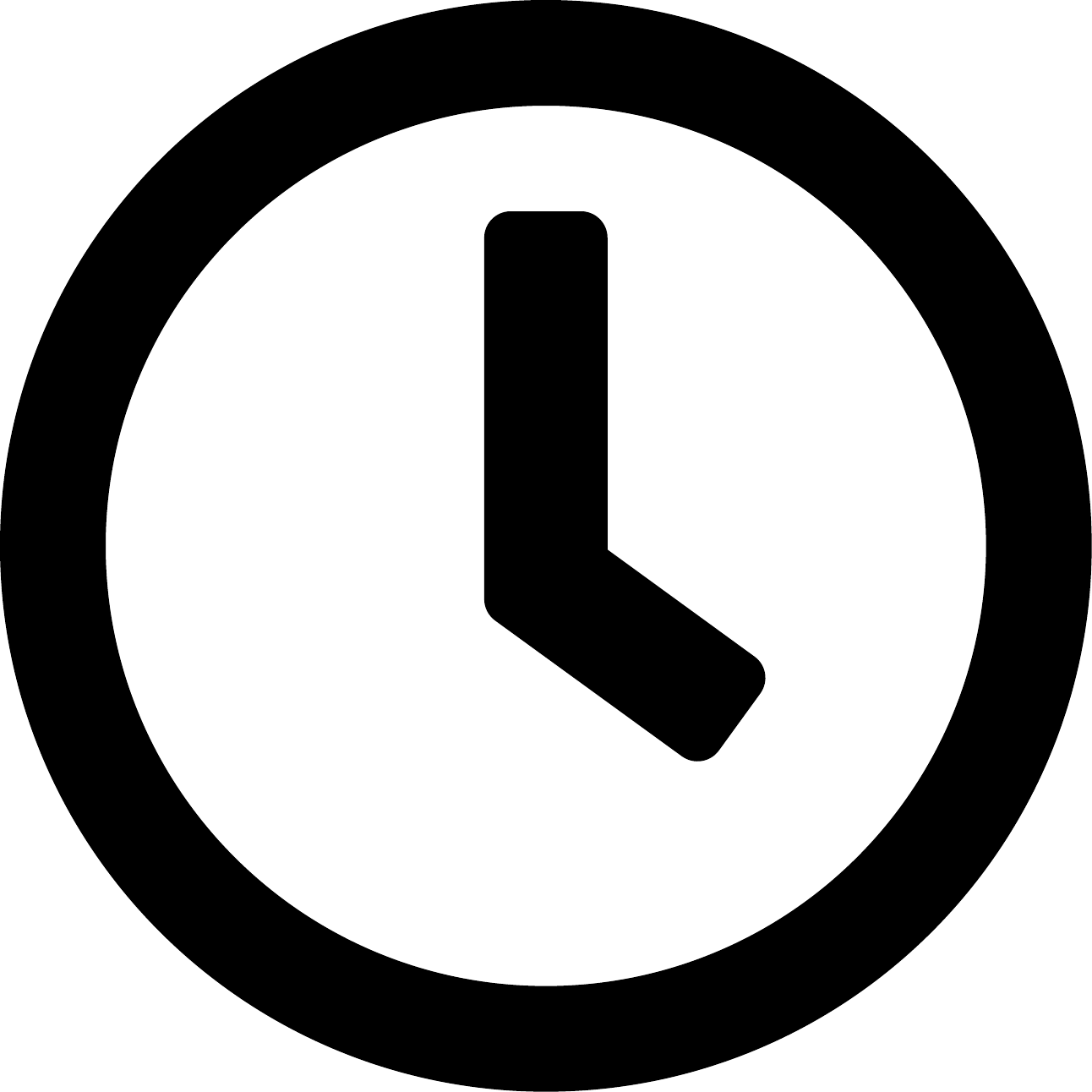}{corpustimevalue}{\TimeTracingPipelineCorpus}{corpus};
  \rstdtpt{corpuspackageicon}{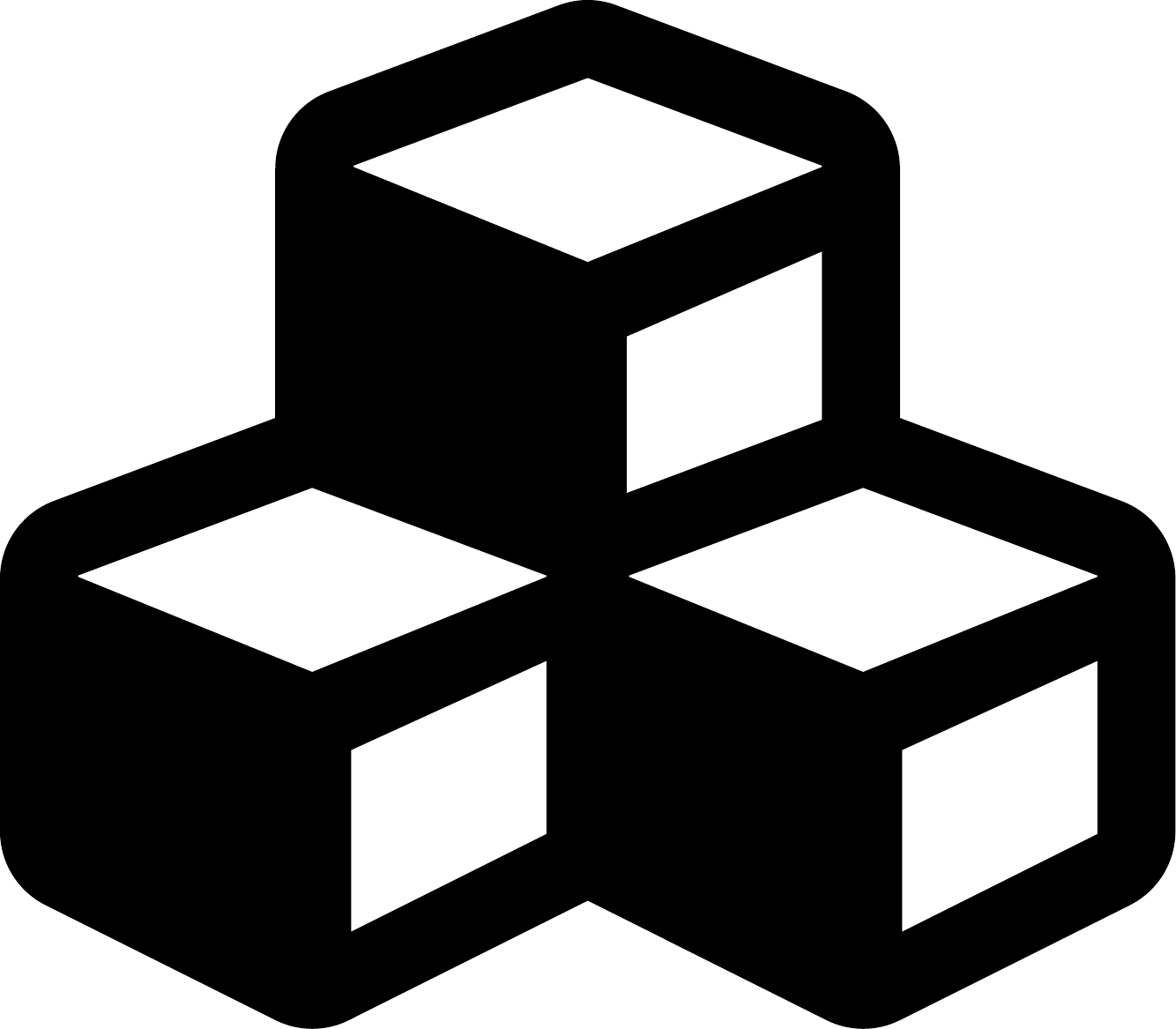}{corpuspackagevalue}{\CountTracingPipelineCorpusPackage}{corpustimeicon};
  \rstdtpt{corpusscripticon}{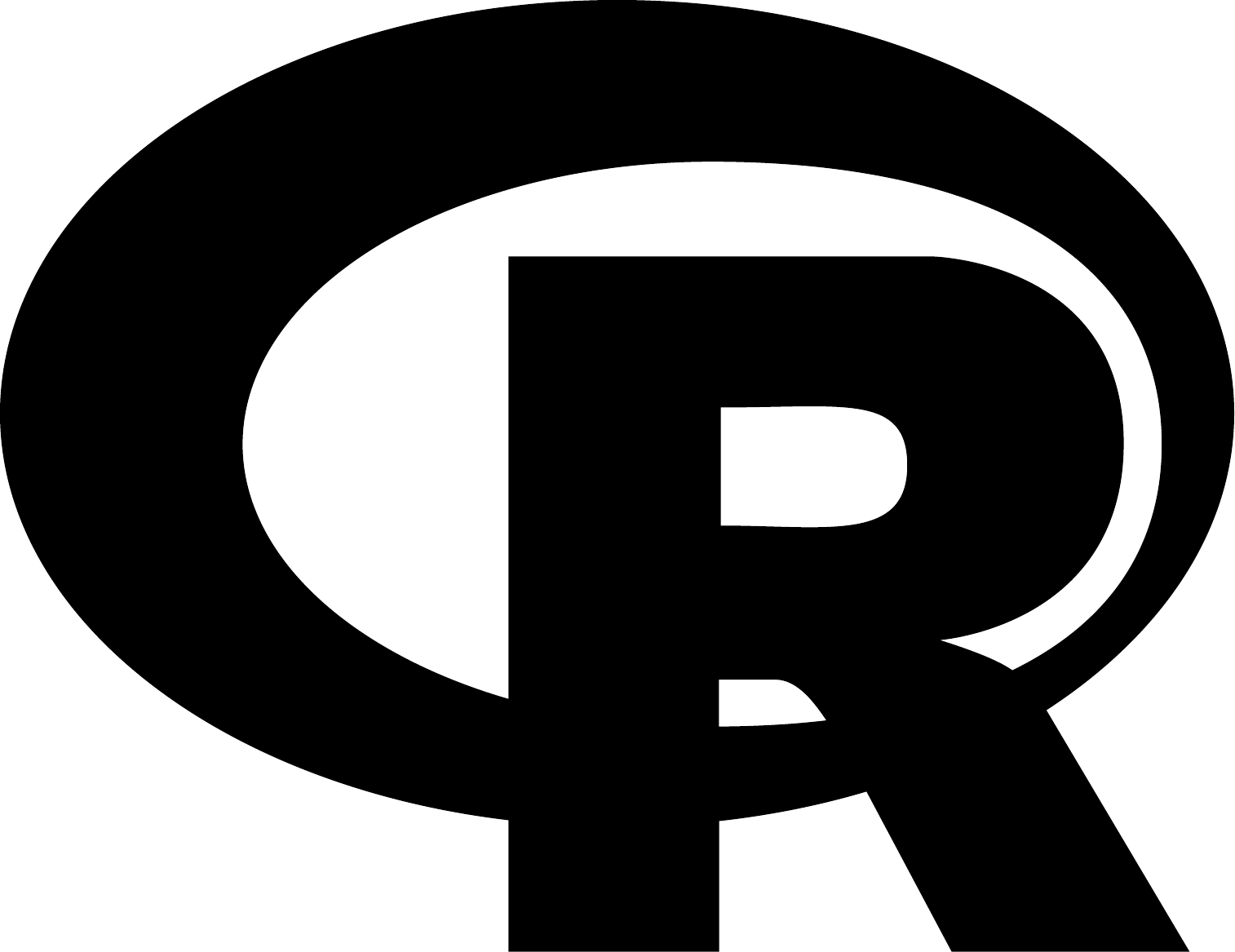}{corpusscriptvalue}{\CountTracingPipelineCorpusScript}{corpuspackageicon};

  \rststage{trace}{Trace}{corpus}{tracebg}{traceol}
  \fstdtpt{tracetimeicon}{clock-regular}{tracetimevalue}{\TimeTracingPipelineTrace}{trace};
  \rstdtpt{tracefileicon}{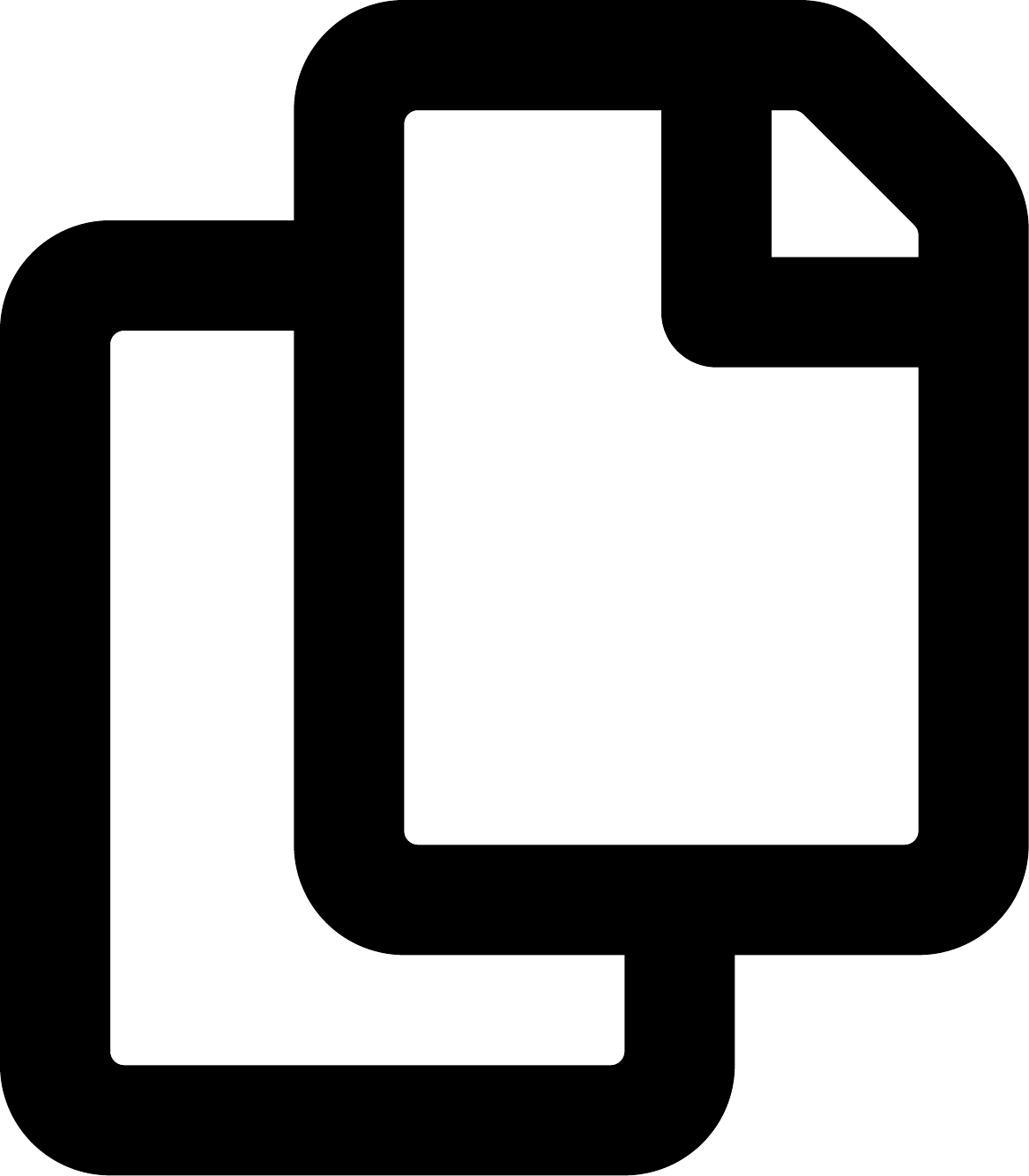}{tracefilevalue}{\CountTracingPipelineTraceFile}{tracetimeicon};
  \rstdtpt{tracesizeicon}{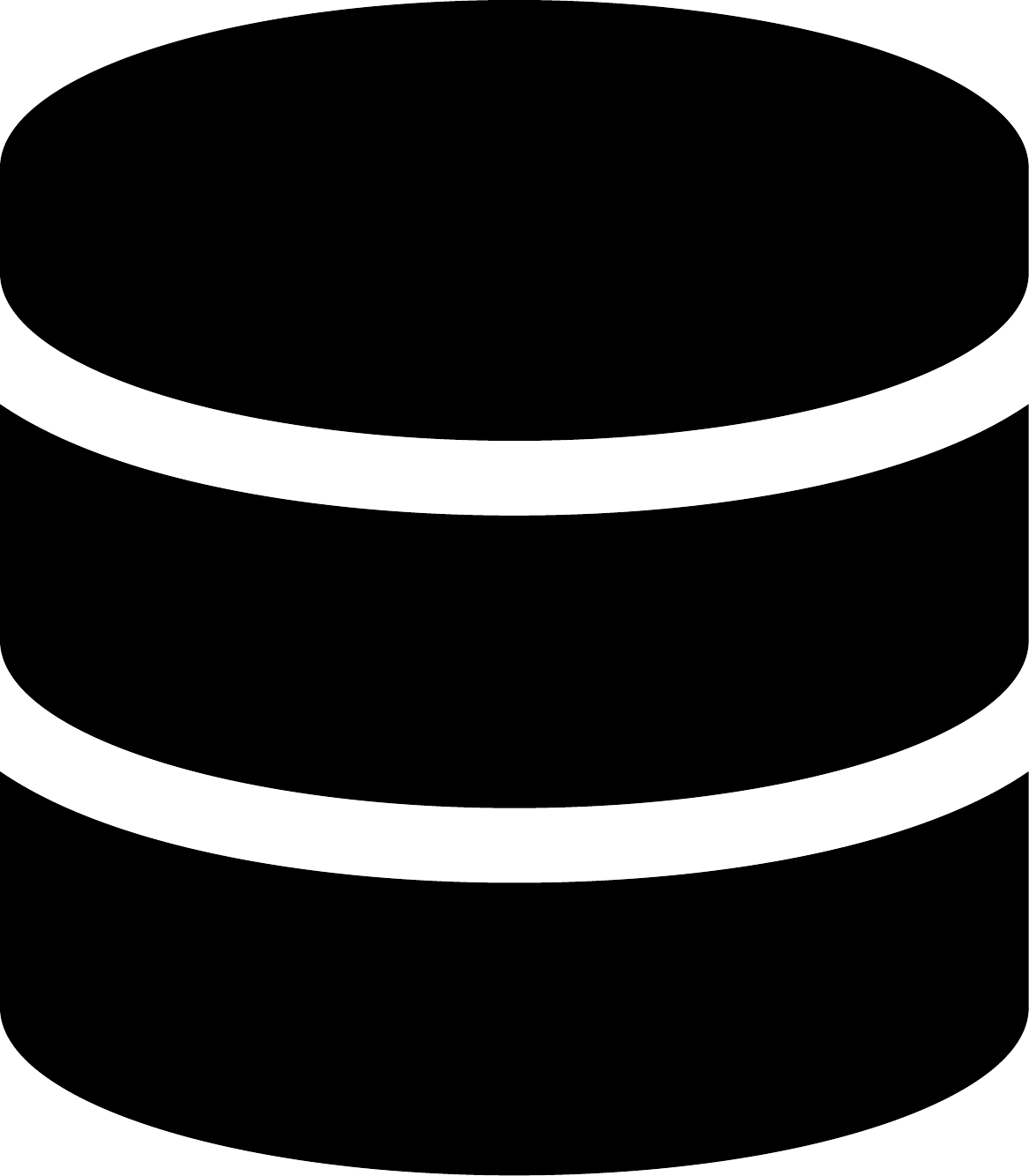}{tracesizevalue}{\SizeTracingPipelineTraceFile}{tracefileicon};

  \rststage{reduce}{Reduce}{trace}{reducebg}{reduceol}
  \fstdtpt{reducetimeicon}{clock-regular}{reducetimevalue}{\TimeTracingPipelineReduce}{reduce};
  \rstdtpt{reducefileicon}{copy-regular}{reducefilevalue}{\CountTracingPipelineReduceFile}{reducetimeicon};
  \rstdtpt{reducesizeicon}{database-solid}{reducesizevalue}{\SizeTracingPipelineReduceFile}{reducefileicon};

  \rststage{combine}{Combine}{reduce}{combinebg}{combineol}
  \fstdtpt{combinetimeicon}{clock-regular}{combinetimevalue}{\TimeTracingPipelineCombine}{combine};
  \rstdtpt{combinefileicon}{copy-regular}{combinefilevalue}{\CountTracingPipelineCombineFile}{combinetimeicon};
  \rstdtpt{combinesizeicon}{database-solid}{combinesizevalue}{\SizeTracingPipelineCombineFile}{combinefileicon};

  \rststage{merge}{Merge}{combine}{mergebg}{mergeol}
  \fstdtpt{mergetimeicon}{clock-regular}{mergetimevalue}{\TimeTracingPipelineMerge}{merge};
  \rstdtpt{mergefileicon}{copy-regular}{mergefilevalue}{\CountTracingPipelineMergeFile}{mergetimeicon};
  \rstdtpt{mergesizeicon}{database-solid}{mergesizevalue}{\SizeTracingPipelineMergeFile}{mergefileicon};

  \rststage{summarize}{Summarize}{merge}{summarizebg}{summarizeol}
  \fstdtpt{summarizetimeicon}{clock-regular}{summarizetimevalue}{\TimeTracingPipelineSummarize}{summarize};
  \rstdtpt{summarizefileicon}{copy-regular}{summarizefilevalue}{\CountTracingPipelineSummarizeFile}{summarizetimeicon};
  \rstdtpt{summarizesizeicon}{database-solid}{summarizesizevalue}{\SizeTracingPipelineSummarizeFile}{summarizefileicon};

  \rststage{report}{Report}{summarize}{reportbg}{reportol}
  \fstdtpt{reporttimeicon}{clock-regular}{reporttimevalue}{\TimeTracingPipelineReport}{report};
  \rstdtpt{reportcharticon}{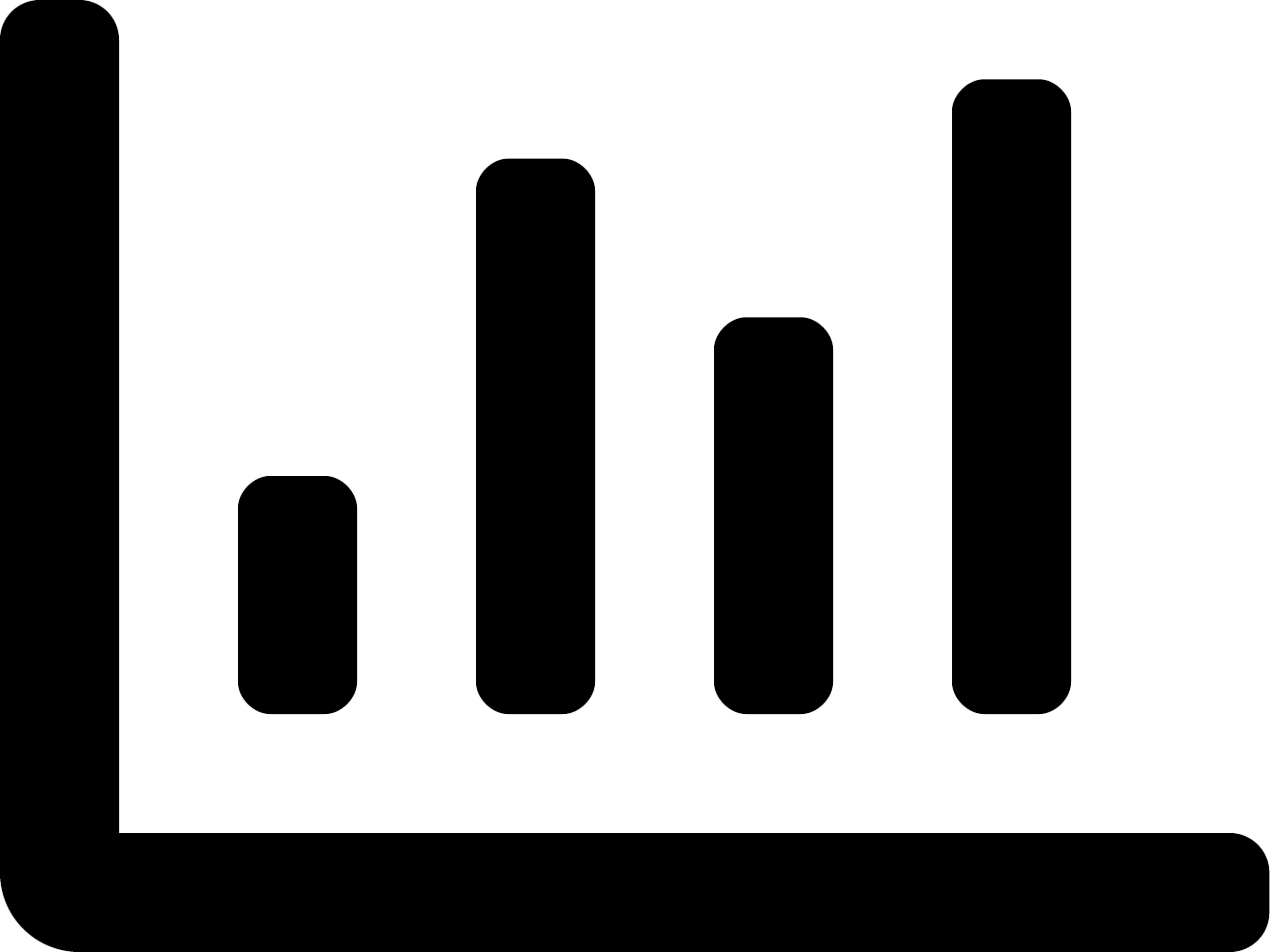}{reportchartvalue}{\CountTracingPipelineReportGraph}{reporttimeicon};
  {
    \node [inner sep = 0, align = center, below = 3.5mm of reportcharticon, minimum width = 0.03 \textwidth] (reportmacroicon) {
      {\scriptsize \LaTeX}
    };
    \node [inner sep = 0, right = 0mm of reportmacroicon, below = 4.3mm of reportchartvalue, minimum width = 0.07 \textwidth, text width = 0.07 \textwidth, align = right] (reportmacrovalue) {\scriptsize \CountTracingPipelineReportMacro};
  };
 \end{tikzpicture}
\caption{Tracing Pipeline}\label{tracerpipeline}
\end{figure}

\noindent
Figure~\ref{tracerpipeline} shows the main stages of our pipeline. We give
corpus size, number of files generated, size of data and time taken by each
step. The remainder of this section details the various stages.

\subsection{Instrumented R}

The instrumented R Virtual Machine is based on GNU-R version \GNURVERSION.
Its goal is to produce program execution traces with all the events
required to answer our research questions. On the face of it, this is not a
difficult task. And, in the end, we only need to add \CountRDyntraceCodeDiff{}
lines of C code to expose an event data structure with fields to describe a variety of
execution events that capture the internal interpreter state.  The challenge
was identifying where to insert those \CountRDyntraceCodeDiff{} lines in an interpreter whose code
is \CountGnuRLoc{} LOC written over twenty-five years by many developers and outside
contributors. The system has grown in complexity with an eclectic mix of
ad hoc features designed to support growing user requirements. For instance,
the code to manage environments and variable bindings in \code{main/envir.c}
is over \CountEnvirLoc{} LOC with \CountEnvirFunDef functions with a large
number of identical code fragments for managing these data structures
duplicated in various files. We succeeded by a lengthy trial and error
process. The events recorded by the instrumented virtual machine are:

\begin{itemize}
\item {\bf\code{Call}, \code{Return}}: at each function call, records
  function's type, arguments, environment and return value.
\item {\bf \code{S3Dispatch}}: at each S3 dispatch, records method name and
  first argument.
\item {\bf \code{S4Dispatch}}: at each S4 dispatch, records method name,
  definition, environment and dispatch arguments.
\item {\bf \code{Eval}}: at each \code{eval}, records the evaluated
  expression and its environment.
\item{\bf \code{Substitute}}: at each \code{substitute}, records the
  arguments.
\item {\bf \code{ArgListEnter}, \code{ArgListExit}}: records expressions that
  are being promised.
\item {\bf \code{CtxtEnter}, \code{CtxtExit}}: records the address of a
  stack frame.
\item {\bf \code{CtxtJmp}}: records the popped stack frames at each non-local return.
\item {\bf \code{PromEnter}, \code{PromExit}}: records evaluated promise and
  when evaluation terminates.
\item {\bf \code{PromRead}}: when a promise's value is read,
  records the promise and its value.
\item {\bf \code{PromSubst}}: generated when a promise's expression is read.
\item {\bf \code{FunLoadStart}, \code{FunLoadEnd}}: generated  when looking
  up a function.
\item {\bf \code{GC}}: generated at each garbage collection cycle.
\item {\bf \code{Alloc}}: generated when memory is allocated.
\item {\bf \code{Free}}: generated memory is reclaimed.
\item{\bf \code{Deserialize}}: generated when an object is  deserialized.
\item {\bf \code{VarDef}}: when a variable is defined, record the
  symbol, value and environment.
\item {\bf \code{VarWrite}}: when a variable is updated, record a
  the symbol, value and environment.
\item {\bf \code{VarRem}}: when a variable is deleted, records the
  symbol and environment.
\item {\bf \code{VarRead}}: when a variable is read, record the
  symbol, value and environment.
\end{itemize}

\noindent
Events can be disabled to ignore implementation details of the virtual
machine and also to avoid recursion. R objects captured in events are
protected from the garbage collector to prevent them from being reclaimed
during analysis.  Table~\ref{events} shows the number of times events
are triggered.

\begin{table}[!h]
  \footnotesize{}
  \caption{Events}
  \label{events}
  \begin{tabular}{lr}
    \hline
    {\bf Call}          & \CountEventCall{}           \\
    {\bf Return}        & \CountEventReturn{}         \\
    {\bf GC}            & \CountEventGC{}             \\
    {\bf Alloc}         & \CountEventAlloc{}          \\
    {\bf Free}          & \CountEventFree{}           \\
    \hline
  \end{tabular}
  \begin{tabular}{lr}
    \hline
    {\bf CtxtEnter}     & \CountEventCtxtEnter{}      \\
    {\bf CtxtExit}      & \CountEventCtxtExit{}       \\
    {\bf CtxtJmp}       & \CountEventCtxtJmp{}        \\
    {\bf S3Dispatch}    & \CountEventSThreeDispatch{} \\
    {\bf S4Dispatch}    & \CountEventSFourDispatch{}  \\
    \hline
  \end{tabular}
  \begin{tabular}{lr}
    \hline
    {\bf Eval}          & \CountEventEval{}           \\
    {\bf ArgListEnter}  & \CountEventArgListEnter{}   \\
    {\bf ArgListExit}   & \CountEventArgListExit{}    \\
    {\bf FunLoadStart} & \CountEventCtxtSensEnter{}  \\
    {\bf FunLoadEnd}  & \CountEventCtxtSensExit{}   \\
    \hline
  \end{tabular}
  \begin{tabular}{lr}
    \hline
    {\bf PromEnter}     & \CountEventPromEnter{}      \\
    {\bf PromExit}      & \CountEventPromExit{}       \\
    {\bf PromRead}      & \CountEventPromRead{}       \\
    {\bf PromSubst}     & \CountEventPromSubst{}      \\
    {\bf Deserialize}   & \CountEventDeserialize{}    \\
    \hline
  \end{tabular}
  \begin{tabular}{lr}
    \hline
    {\bf VarDef}        & \CountEventVarDef{}         \\
    {\bf VarWrite}      & \CountEventVarWrite{}       \\
    {\bf VarRem}        & \CountEventVarRem{}         \\
    {\bf VarRead}       & \CountEventVarRead{}        \\
    {\bf Substitute}    & \CountEventSubstitute{}     \\
    \hline
  \end{tabular}
\end{table}

\subsection{Tracer}
The tracer is a small R package (73 LOC) that calls into a larger C++
library (6,080 LOC).  It is loaded in the instrumented R virtual machine
and, during program execution, it maintains objects that model various
aspects of the program such as functions, calls, promises, variables,
environments, stacks and stack frames. As events are generated, the tracer
updates its model of the state.
The tracer is able to
process \CountTotEventRate events per second on our benchmark machine.

\noindent
Some design decisions allowed the tracer to scale.  Firstly, copying model
objects is avoided as much as possible. They are created by a singleton
factory that caches them in a global table. This optimization pays off as
model objects are large and costly to copy.  But keeping these objects alive
too long will increase footprint and hinder any attempt at running multiple
tracers on the same machine in parallel. To reduce tracing footprint, the R
garbage collector was modified so that model objects can be deallocated as
soon as the R object they represent is freed. One slightly surprising design
choice is to link all model objects together. This pays off when an event
triggers a cascade of changes to model objects. This comes at a price of
course, as lists of model objects are circular it is necessary to perform
reference counting to reclaim them. One last implementation trick is the use
of a shadow stack that mirrors the stack maintained by the R virtual
machine. The shadow stack is used to look up data after a \code{longjump}.

The tracer generates large amounts of data. Our first prototype used Sqlite
to store the generated data. However, we found the approach limiting. During
development we kept running into errors because the database schema and the
tracer were out of sync.  Due to the iterative nature of data analysis, we
were modifying the schema frequently and this became a pain point.
Furthermore, our database was normalized, thus requiring join operations in
the analysis. At our scale, these joins were expensive, causing database
operations to run for days. Lastly, insertions ended being a bottleneck; we
could trace fewer than 1000 packages per day and filled up a 1 TB disk.  In
the end, we implemented a custom format. As the event stream has substantial
amounts of redundancy, we applied streaming compression on the fly.
Compression yields an average 10x saving in space and 12x improvement in
loading time.

\subsection{Execution}

For each package to be analyzed our infrastructure must extract executable
code from that package. Extraction invokes an R API which locates executable
code snippets in the documentation and RMarkdown files. Files in the test
directory are copied as is. All snippets and tests set up the tracer and
initialize it with paths to input and output data before execution. For each
package, the tracer generates 12 data files and 4 status files. These files
denote the different possible states of the tracing. They allow the
infrastructure to discard data from failing programs. The R scripts
responsible for generating traces do extensive logging of intermediate steps
for debugging purposes.

\subsection{Post-processing}

This part of the pipeline analyzes the raw data. It is 4K lines of R code.
Scale was our major challenge. We faced difficulties both due to execution
time and data size.  In the \ProgramCount programs that were traced, the
tracer observed \CountExpression expression evaluations, \CountCall
calls to \CountFunction functions and \CountProm promises.  The raw data
generated by the tracer is \SizeTracerRawData{} but the reduced data is just
\SizeReducedData{}. In hindsight, it appears that incorporating analysis in the
tracer, i.e.,
pre-summarizing data in C++ would have been beneficial.  However, this would
require knowing ahead of time all the analysis that we would perform.  Part
of the challenge lies in the fact that the event of interest and attached
analyses were not fixed ahead of time.  Pre-summarized data makes it harder
to pose new questions. It also makes it harder to detect bugs because
summarized data resists correlation with actual code.

The pipeline steps are detailed next.
\begin{itemize}
\item {\it Prescan}: Scan the raw data directory and output a list of all the
subdirectories that contain raw data. There is a directory per package and
multiple files in each directory.
\item {\it Reduce}: Given a list of directories, this step uses GNU Parallel
  to partially summarize the raw data.  This is the most expensive step in
  terms of size and speed. Since the data files are large, we must limit the
  degree of parallelism drastically to avoid running out of memory.
\item {\it Scan}: Create a list of all the files successfully reduced.
\item {\it Combine}: Combine information from all the programs into a single
  data table per analysis question.
\item {\it Summarize}: Compute summaries of the merged data for:
(1) event frequency, (2) object frequency, (3) functions with their
definitions, (4) argument information, (5) escaped arguments, (6)
information about parameters, (7) information about promises.
\item {\it Report}: Generate graphs and tables from an RMarkdown notebook as
  well as \LaTeX\xspace macros for inclusion in the paper.
\end{itemize}

\subsection{Threats to Validity}

We have mentioned in the introduction that code coverage is a worry for any
dynamic approach. There are two additional points to consider. Firstly, C and C++
functions can bypass the R extension API and directly modify R objects'
internals. For example, set a promise's value without going through the API
thus obviating our hooks. Such behavior breaks the R semantics and is error
prone as the R internals do change. We have not observed this behavior in
practice, but given the large number of packages, it may happen.
Secondly, we disable the bytecode compiler for this study. Since the compiler
eliminates promises for literal arguments, we observe more promises than
actually created by the \GNUR Virtual Machine under its default settings.

\newpage
\section{Corpus of R Programs}

The corpus used in this study was assembled on August 1st, 2019 from the two
main code repositories, namely the \emph{Comprehensive R Archive Network}
(CRAN)~\cite{LIgges2017} and \emph{Bioconductor}~\cite{bioc}.  Both are
curated repositories; to be admitted packages must conform to some
well-formedness rules. In particular, they must contain use-cases and tests
along with the data needed to run them. We believe this corpus is
representative of sophisticated uses of the R language.  Anecdotal evidence
suggests that the majority of R code written is made up of small scripts,
straight-line sequences of package calls, that read data, apply some models
to it and then visualize the results.  Most end-users neither define
functions nor write loops, their code is simple. Without a source of
end-user code it is not possible to validate this hypothesis, but if true
then our corpus is representative of the interesting R code.  R is also used
in many industrial settings that do not publish their code to open source
repositories. We have no information on those use-cases.

Our snapshot of CRAN includes \CntCranPkg packages, and for Bioconductor,
\CntBiocPkg packages. Bioconductor is also used to store data,
\CntBiocSoftPkg packages contain software, \CntBiocDataPkg contain data and
\CntBiocWflPkg are so-called workflows. Starting with \CntTotalPkg software
packages, our scripts downloaded and successfully installed \CntInstPkg of
them. The reasons some packages did not install were varied, they included
missing dependencies and compiler errors.

\begin{wraptable}{r}{7cm}
  \vspace{-6mm}
  \small
  \caption{Corpus}\label{corpus}
  \begin{tabular}{l|rrrl} 
&\bf Tests&\bf Examples&\bf Vignettes&\\\hline
\multirow{2}{*}
{Scripts}&\CountInstalledTestsRScripts&\CountInstalledExamplesRScripts&\CountInstalledVignettesRScripts&\it Install\\
&\CountTracedTestsRScripts&\CountTracedExamplesRScripts&\CountTracedVignettesRScripts&\it Trace\\\hline
\multirow{2}{*}
{LOC}&\CountInstalledTestsRCode&\CountInstalledExamplesRCode&\CountInstalledVignettesRCode&\it Install \\
&\CountTracedTestsRCode&\CountTracedExamplesRCode&\CountTracedVignettesRCode&\it Trace\\\hline
\end{tabular}
\end{wraptable}

These install errors may be fixable but automating those fixes would be
hard. We chose to discard the packages that could not be installed.  Out of
the installed packages, we were able to successfully record execution traces
for \CountCorpusPackages packages.  Some packages did not trace owing to
run-time failures.  Again, we discard the failing packages on the grounds of
having a sufficient number of running ones

For each package, our scripts gathered runnable code from three different
sources: test cases, examples and vignettes. Test cases are typically unit
tests written to exercise individual functions, while examples and vignettes
demonstrate the expected end-user usage of the particular package.  These
use-cases may load other packages and access data shipped with the package
or obtained from the internet.  Table~\ref{corpus} gives the number of
scripts of each kind that could be installed and the number of scripts that
were successfully traced.  In terms of lines of code, we exercised
\CountTracedSourceRCode lines of R and \CountTracedSourceCCode lines of C.
The total size of our database after analysis is \SizeTotalTotal.

We observed \CountCall calls to \CountFunction functions.  
\PercBuiltinCall of the calls are made to \CountBuiltinFunction builtin
functions, and
\PercSpecialCall are made to \CountSpecialFunction special functions.
The remainder are calls to R functions. Of these,
\PercThreeCall are calls to
\CountThreeFunction different S3 methods and
\PercFourCall are calls to \CountFourFunction S4 methods.  There are
\CountOrdFunction plain R functions.

The number of times \code{delayedAssign} is called is
\CountDelayedAssignCall, \code{force} is called \CountForceCall times,
\code{forceAndCall} is invoked \CountForceAndCallCall times and
\code{substitute} \CountSubstituteCall times.  Functions exit due to
lonjumps \CountJumpedCall times (this marks explicit use of the
\code{return} function).

\begin{figure}[H]
  \includegraphics[width=1.05\textwidth]{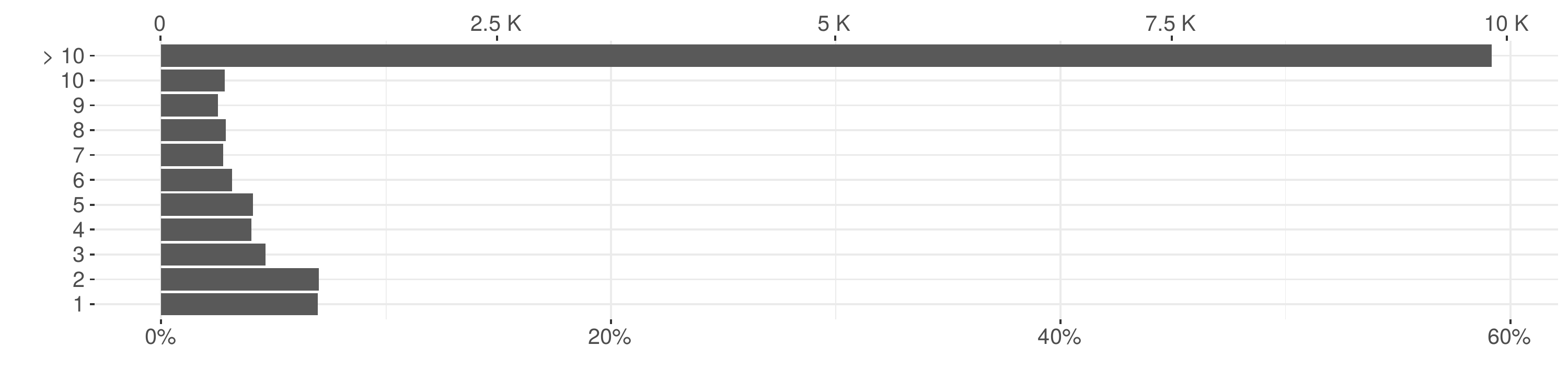}
  \vspace{-1cm}\caption{Functions per package} \label{p}
\end{figure}

Figure~\ref{p} shows how many functions were exercised in each test
package. More than 9 K (\PercManyFunctionPackages{}) of the packages had over
10 functions called. On the other hand 2.7 K of the packages had a single
function invoked (\PercSingleFunctionPackages{}). These may either be small
packages, or, more likely, the provided tests have low coverage.

\begin{figure}[H]
  \includegraphics[width=1.05\textwidth]{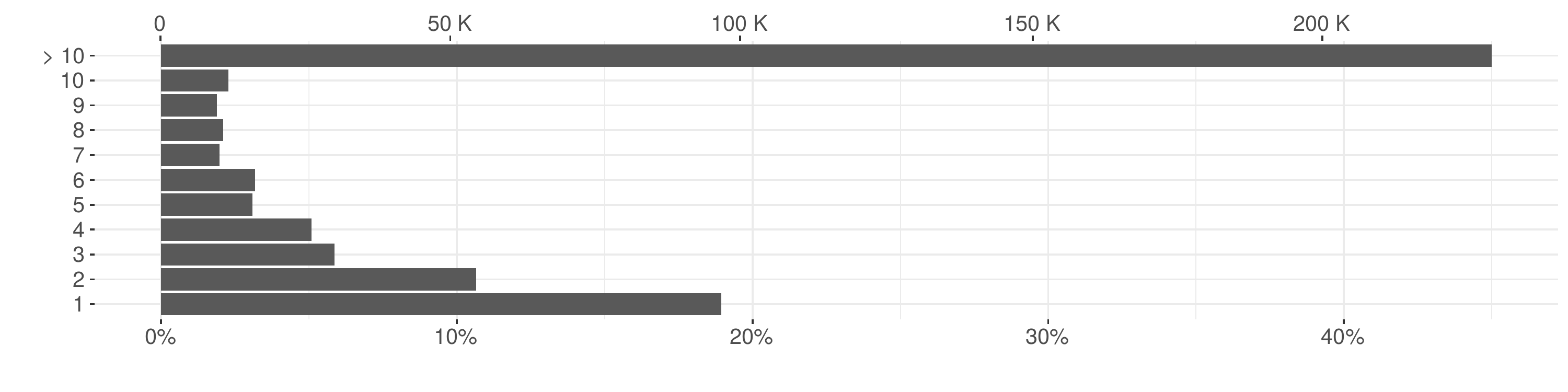}
  \vspace{-1cm}\caption{Function calls} \label{figure:closure_calls}
\end{figure}

\noindent
Figure~\ref{figure:closure_calls} shows how many times functions were
called.  \PercManyCallClosures of the exercised functions were called more
than ten times and \PercSingleCallClosures of the functions were called only
once.  Clearly functions that are called only once may lack coverage.

\begin{figure}[H]
  \includegraphics[width=1.05\textwidth]{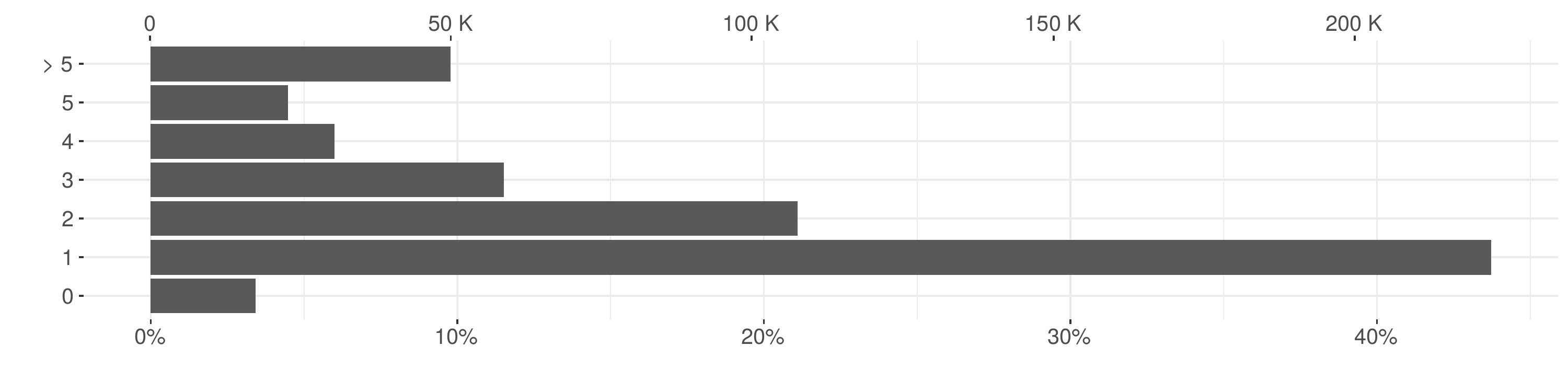}
  \vspace{-1cm}\caption{Formal Parameters} \label{figure:closure_parameter_distribution}
\end{figure}

\noindent
Figure~\ref{figure:closure_parameter_distribution} shows the number of
parameters per function; \PercSingleParameterClosures have only one, while
\PercManyParameterClosures have more than five.  The function
\CodeMaximumParameterFunctionName has the most with
\CountMaximumParameterCount parameters. There are \CountParameter distinct
parameters.

\noindent
Of the \CountArgument arguments that were passed at run-time, \PercDefArgument
were default arguments, \PercNonDefArgument were non-default arguments and the
remaining \PercMissingArgument were missing arguments.

\newpage
\section{Analyzing Laziness Usage Patterns}

This section presents the results of our empirical study of call-by-need in
the R language.

\subsection{Life Cycle of Promises}

The first research question we address is how promises are created and used
in the wild.

\begin{center}
{\bf RQ1}: \emph{What is the life cycle of R promises in the corpus?}
\end{center}

\begin{figure}[!h]
\includegraphics[width=1.05\textwidth]{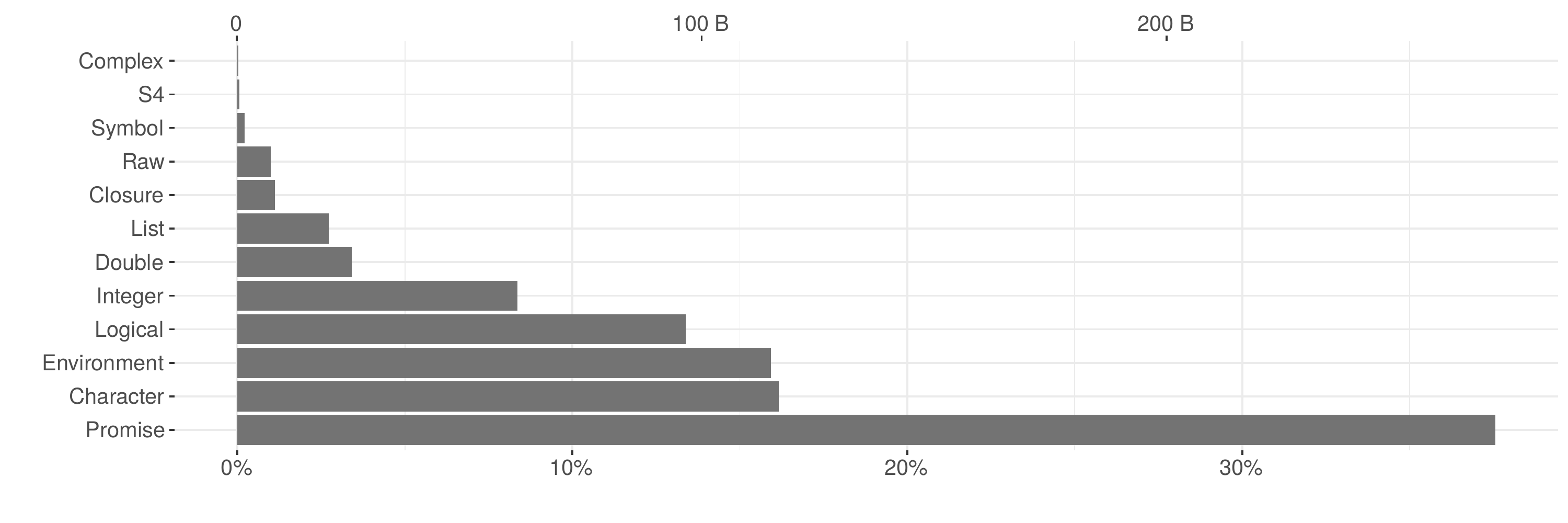}
\vspace{-1cm}\caption{Object counts} \label{figure:object-count-by-type}
\end{figure}

\noindent
In our corpus, and likely, all R programs, promises are the most frequently
allocated object. We observed the creation of \CountProm promises.  For
context, Figure~\ref{figure:object-count-by-type} shows the distribution of
application-level objects; most are vector of characters, logical, integers,
and doubles, in that order. Lists are often used in package code and
internal functions. Raw values hold uninterpreted byte strings. Closures
represent functions and environments map names to value, symbols are
language-level names and S4 are instance of classes.  Environments are
frequently observed because one is created for each function call, but they
are also, albeit rarely, used as hashmaps in user code.

\paragraph{Where are promises created?} 
Argument lists account for \PercArgProm of promises. The remainder are used
for lazy loading of functions, are created by calls to \c{delayedAssign}, or
in internal functions of the R virtual machine.  One could expect that there
would be more promises than values since every operation in R is a function
call. This is not the case. Some values are composite of multiple simpler
elements (e.g. data frames) and these are wrapped in a single
promise. Values can be returned without being bound to a promise. Lastly,
calls to internal (builtin and special) functions do not pack arguments in promises.

\begin{figure}[!h] \centering
\includegraphics[width = \textwidth]{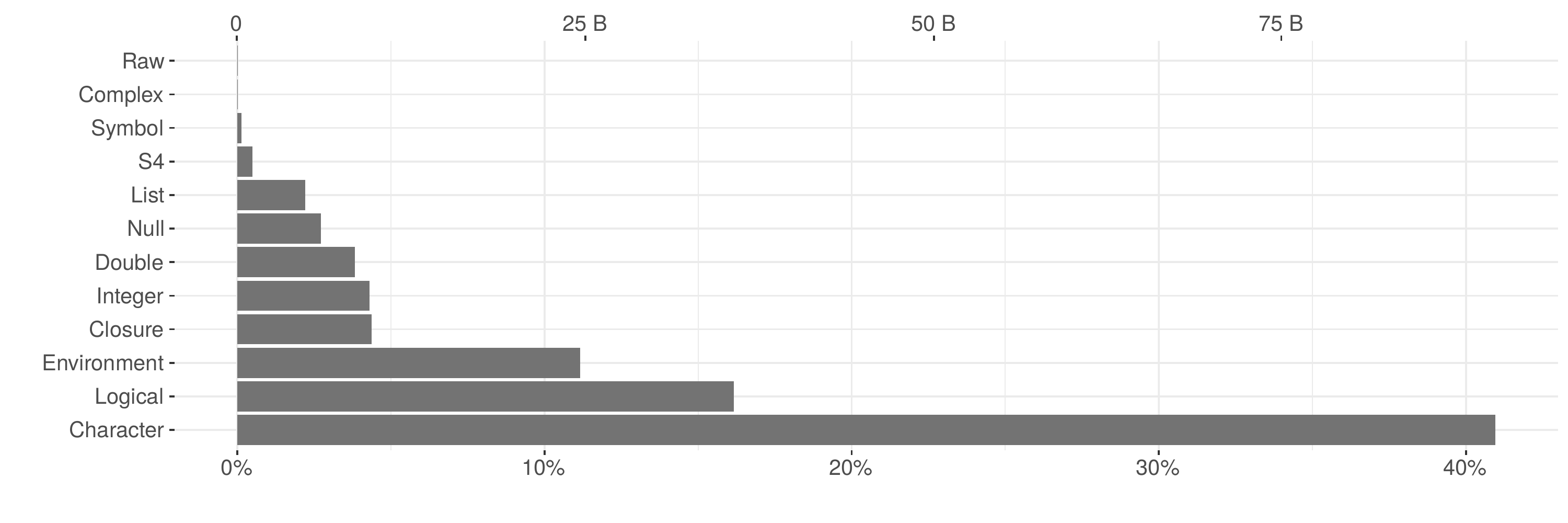}\vspace{-.8cm}
\caption{Promise results} \label{prom-type}
\end{figure}

\paragraph{What do promises yield?}
Figure~\ref{prom-type} shows the contents of promises that were forced. The
most common types are character vector, logical vector, environment, closure,
integer vector, double vector, null and list, in that order. The presence of
null values is explained by the fact that many default parameter values are set
to null and these default values are promised. S4 objects are rarely used in R
programs outside of Bioconductor packages. Symbol, complex vector and raw
vector are also quite rare in practice.

\newpage

\begin{figure}[!h] \centering
\includegraphics[width = \textwidth]{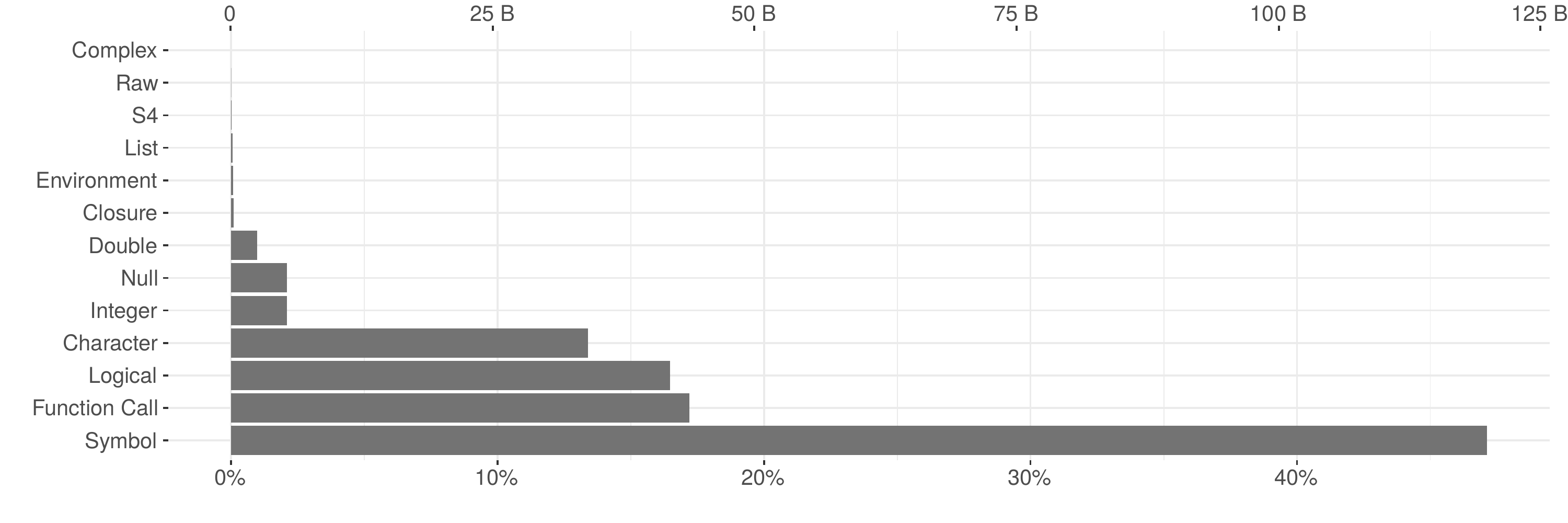}\vspace{-.51cm}
\caption{Promise expressions} \label{prom-exp-type}
\end{figure}

\noindent Figure~\ref{prom-exp-type} shows the content of the expression
slot of promises, i.e. their code.  Only
\PercArgumentPromiseExpressionFunctionCall of promises contain a function
call, e.g. \code{1+2} or \code{f(z)}. The majority contain a single symbol
to be looked up in the promise's defining environment, e.g. \code{x}.  That
symbol may be bound to another promise, in which case forcing the promise
will be recursive.  Promises can also hold inlined scalar constants, such as
a single double \c{1.1}.

\begin{figure}[!h]\includegraphics[width=\textwidth]{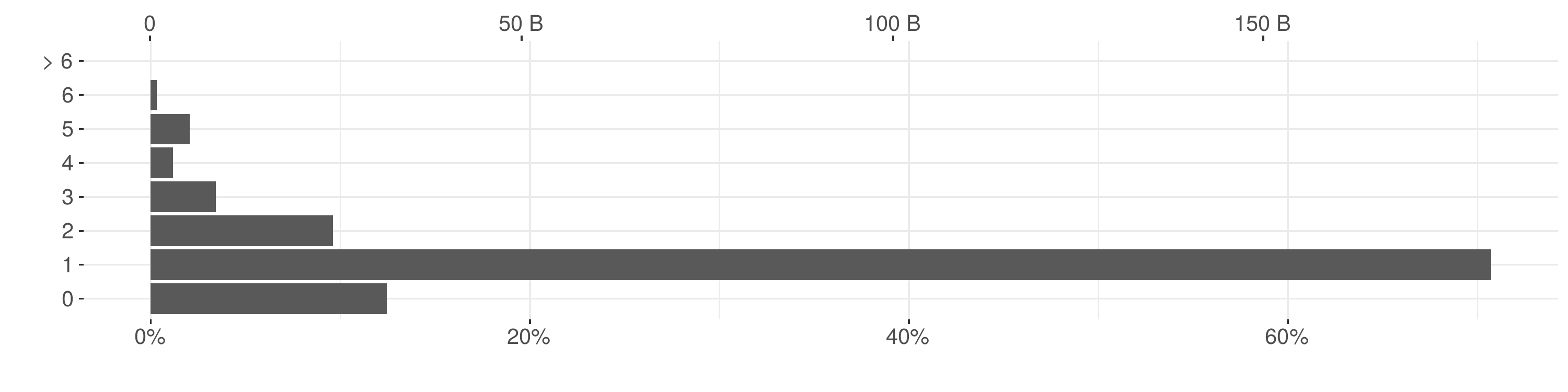}
\vspace{-8mm}\caption{Reads}\label{promreads}
\end{figure}

\paragraph{How often are promises accessed?}
\PercForceProm of argument promises are forced; the remaining went unused.
Unused arguments are not unusual, some functions have over twenty
parameters, and many of these are only needed in special
circumstances. Figure~\ref{promreads} shows, for each individual promise,
the number of times its value was read.  Most promises are used once
(forced), \PercTwo are accessed twice, and \PercThree are accessed three
times.

\begin{figure}[!h]\includegraphics[width=\textwidth]{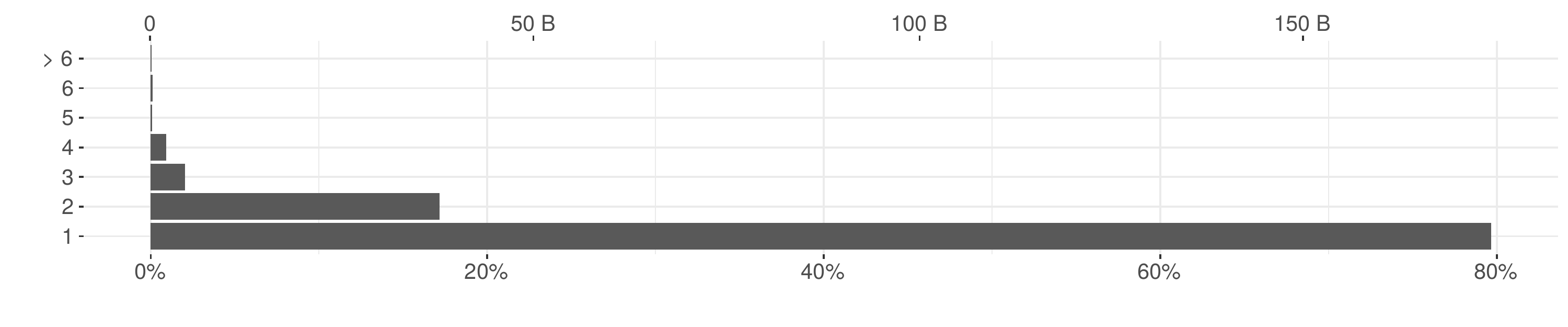}
\vspace{-1cm}\caption{Force depth} \label{forcedepth}
\end{figure}

\paragraph{How far are promises forced from their creation?}
Promises can be passed from one function to the next, traveling down the
call stack. Regardless of the distance from the frame that created them,
promises evaluate in their creation environment. But the farther from
creation, the harder it is for a compiler to optimize them.
Figure~\ref{forcedepth} shows the distance between promise creation and forcing.
\PercPromiseZeroDepth promises are evaluated in the callee,
\PercPromiseOneDepth promises are forced two level down, and the remaining
\PercPromiseMultipleDepth are evaluated deeper.

\paragraph{How long do promises live?}
Promises are short-lived, over \PercSingleGCPromises do not survive one
garbage collection cycle. This confirms the folklore that most objects die
young.  It also means that promises are exerting pressure on the memory
subsystem of the virtual machine. Only \PercMultipleGcPromises of promises
survive multiple cycles.  Of those, \PercMultipleGcNonArgumentPromises are
non-argument promises and \PercMultipleGcEscapedArgumentPromises are escaped
arguments. As mentioned earlier, non-argument promises are created
explicitly through \code{delayedAssign} and implicitly through lazy-loading
of package code; both are expected to be long-lived. Escaped promises are
promises that outlive their defining function. They may be long-lived as
well. Of the long-lived promises, \PercMultipleGcUnescapedArgumentPromises
are argument promises; their longevity is likely due to long running
functions.

\begin{wraptable}{r}{7.7cm}
  \caption{Promise life cycle} \label{lifecycle}
  \begin{tabular}{lr} 
    \hline
    \TextLifecycleUnescapedArgumentsa{}&\PercLifecycleUnescapedArgumentsa{}\\
    --&\PercLifecycleUnescapedArgumentsb{}\\
    \TextLifecycleUnescapedArgumentsc{}&\PercLifecycleUnescapedArgumentsc{}\\
    \TextLifecycleUnescapedArgumentsd{}&\PercLifecycleUnescapedArgumentsd{}\\
    \TextLifecycleUnescapedArgumentse{}&\PercLifecycleUnescapedArgumentse{}\\
    \TextLifecycleUnescapedArgumentsf{}&\PercLifecycleUnescapedArgumentsf{}\\
    \TextLifecycleUnescapedArgumentsg{}&\PercLifecycleUnescapedArgumentsg{}\\
    \hline
    \multicolumn{2}{c}{\small{a. Argument}}\\
    \hline
  \end{tabular}\hspace{1mm}
  \begin{tabular}{lr} 
    \hline
    \TextLifecycleEscapedArgumentsa{}&\PercLifecycleEscapedArgumentsa{}\\
    \TextLifecycleEscapedArgumentsb{}&\PercLifecycleEscapedArgumentsb{}\\
    \TextLifecycleEscapedArgumentsc{}&\PercLifecycleEscapedArgumentsc{}\\
    \TextLifecycleEscapedArgumentsd{}&\PercLifecycleEscapedArgumentsd{}\\
    \TextLifecycleEscapedArgumentse{}&\PercLifecycleEscapedArgumentse{}\\
    \TextLifecycleEscapedArgumentsf{}&\PercLifecycleEscapedArgumentsf{}\\
    \TextLifecycleEscapedArgumentsg{}&\PercLifecycleEscapedArgumentsg{}\\
    \hline
    \multicolumn{2}{c}{\small{b. Escaped}}\\
    \hline  \end{tabular}
  \hspace{1mm}
  \begin{tabular}{lr}
    \hline
    \TextLifecycleNonArgumentsa{}&\PercLifecycleNonArgumentsa{}\\
    --&\PercLifecycleNonArgumentsb{}\\
    \TextLifecycleNonArgumentsc{}&\PercLifecycleNonArgumentsc{}\\
    \TextLifecycleNonArgumentsd{}&\PercLifecycleNonArgumentsd{}\\
    \TextLifecycleNonArgumentse{}&\PercLifecycleNonArgumentse{}\\
    \TextLifecycleNonArgumentsf{}&\PercLifecycleNonArgumentsf{}\\
                                 &\\
    \hline
    \multicolumn{2}{c}{\small{c. Non-Argument}}\\
    \hline
  \end{tabular}
\end{wraptable}

\paragraph{What are promise life cycles?}
If we characterize the life of a promise by the events that affect it,
promise life cycles can be summarized by sequences of events. Ignoring
creation and reclamation, the events of interest are {\tt\underline{F}}orce,
{\tt\underline{R}}ead, {\tt\underline{M}}eta-program,
{\tt\underline{E}}scape, {\tt\underline{A}}ssign, and
de{\tt\underline{S}}erialize. Note that forcing a promise is an implicit
read. We observed \CountLifecycleTotal unique life cycles.
Table~\ref{lifecycle} shows the most frequent sequences for (a) argument
promises (\CountLifecycleUnescapedArguments unique sequences), (b) escaped
argument promises (\CountLifecycleEscapedArguments sequences) and (c)
non-argument promises (\CountLifecycleNonArguments sequences).  For argument
promises the two most common sequence are {\tt F} (a promise that is forced)
and the empty sequence (unused promise).  The next sequences are forces
followed by a growing number of reads. Meta-programming occurs only
infrequently. For escaped promises the most frequent sequence is {\tt EF},
the promise escapes and is forced later. The second most frequent sequence
is {\tt FRER}, the promise is forced, read, escapes and is read again.
Lastly, non-argument promises are most often created and assigned a value in
the C code. About 0.2\% of these promises are obtained from deserialization
({\tt S}) owing to lazy-loading of packages.

\newpage

\paragraph{Does context sensitive lookup  force promises?}
Looking up a function name such as \c{f()} may force a promise if that name
is bound to one in the environment. If the promise yields a closure, that
closure will be invoked, otherwise lookup continues.  This allows
``harmless'' shadowing of function names as seen in Figure~\ref{sha} where
\code{addToGList.grob} from package \code{grid} defines a parameter
\code{gList} that shadows a function of the same name defined by the same
package.

\begin{figure}[!h]
\begin{lstlisting}[style=R]
 addToGList.grob <- function(x, gList) 
     if (is.null(gList)) gList(x) else { gList[[length(gList)+1L]]<-x; return(gList) }
\end{lstlisting} 
\caption{Shadowing}\label{sha}\end{figure}

\noindent
We observed \CountConSenArgForce function lookups (out of a total of
\CountCall lookups) which caused a promise to be forced.  Out of those,
\PercConSenArgForceSuccess yielded a closure. The \PercConSenArgForceFailure
that did not yield a closure are cases where a function name was shadowed.
Table~\ref{consens} shows the 30 most commonly shadowed function names and
the number of functions in which shadowing happens.  Many of those names
correspond to common R functions such as \code{c}, \code{names},
\code{print} and \code{max}.

\begin{table}[!h]
  \caption{Context sensitive lookup} \label{consens}
  \begin{tabular}{lr}
    \hline
    \CodeConSenSymbola{}&\CountConSenSymbola{}\\
    \CodeConSenSymbolb{}&\CountConSenSymbolb{}\\
    \CodeConSenSymbolc{}&\CountConSenSymbolc{}\\
    \CodeConSenSymbold{}&\CountConSenSymbold{}\\
    \CodeConSenSymbole{}&\CountConSenSymbole{}\\
    \CodeConSenSymbolf{}&\CountConSenSymbolf{}\\
    \hline
  \end{tabular}
  \quad
  \begin{tabular}{lr}
    \hline
    \CodeConSenSymbolg{}&\CountConSenSymbolg{}\\
    \CodeConSenSymbolh{}&\CountConSenSymbolh{}\\
    \CodeConSenSymboli{}&\CountConSenSymboli{}\\
    \CodeConSenSymbolj{}&\CountConSenSymbolj{}\\
    \CodeConSenSymbolk{}&\CountConSenSymbolk{}\\
    \CodeConSenSymboll{}&\CountConSenSymboll{}\\
    \hline
  \end{tabular}
  \quad
  \begin{tabular}{lr}
    \hline
    \CodeConSenSymbolm{}&\CountConSenSymbolm{}\\
    \CodeConSenSymboln{}&\CountConSenSymboln{}\\
    \CodeConSenSymbolo{}&\CountConSenSymbolo{}\\
    \CodeConSenSymbolp{}&\CountConSenSymbolp{}\\
    \CodeConSenSymbolq{}&\CountConSenSymbolq{}\\
    \CodeConSenSymbolr{}&\CountConSenSymbolr{}\\
    \hline
  \end{tabular}
  \quad
  \begin{tabular}{lr}
    \hline
    \CodeConSenSymbols{}&\CountConSenSymbols{}\\
    \CodeConSenSymbolt{}&\CountConSenSymbolt{}\\
    \CodeConSenSymbolu{}&\CountConSenSymbolu{}\\
    \CodeConSenSymbolv{}&\CountConSenSymbolv{}\\
    \CodeConSenSymbolw{}&\CountConSenSymbolw{}\\
    \CodeConSenSymbolx{}&\CountConSenSymbolx{}\\
    \hline
  \end{tabular}
  \quad
  \begin{tabular}{lr}
    \hline
    \CodeConSenSymboly{}&\CountConSenSymboly{}\\
    \CodeConSenSymbolz{}&\CountConSenSymbolz{}\\
    \CodeConSenSymbolA{}&\CountConSenSymbolA{}\\
    \CodeConSenSymbolB{}&\CountConSenSymbolB{}\\
    \CodeConSenSymbolC{}&\CountConSenSymbolC{}\\
    \CodeConSenSymbolD{}&\CountConSenSymbolD{}\\
    \hline
  \end{tabular}
\end{table}

\paragraph{Do promises force each other?}
A parameter list can include parameters with default values that refer to
each other. This is a semantic quirk that can lead to promises forcing one
another.  Consider Figure~\ref{one} and function \c{sample.int} from the
\c{base} package.  Parameter \code{useH} has a default value that depends
on the other four arguments and \c{s} depends on \code{n}. Function
\c{rmslash} has a recursive dependency between \c{center} and
\c{Scatter}. The function expects at least one of those to be provided at
each call site, if this is not the case an error will be reported.

\begin{figure}[H]
\begin{lstlisting}[style=R]
 sample.int <- function(n,s=n,r=F,p=NULL,useH=(!r && is.null(p) && s<=n/2 && n>1e+07)) 
     if (useH) .Internal(sample2(n,s)) else .Internal(sample(n,s,r,p))

 rmslash <- function(center=rep(0,nrow(Scatter)), Scatter=diag(length(center))) {
     if (length(center) != nrow(Scatter)) stop("<error-message>")
 ...
\end{lstlisting} 
  \caption{Argument lists}\label{one}
\end{figure}

\noindent
We found \CountMutualForcingArgumentPositions default value expressions in
\CountMutualForcingFunctions functions.  At run-time, there were
\CountMutualForcing default expressions forcing other parameters. This is
\PercMutualForcing{} of the forced promises.

\paragraph{Does method dispatch force promises?}
The two widely used object systems, S3 and S4, have an impact on promises.
In order to find which method to invoke, one or more arguments must be
forced.  Overall, only \PercDispatch of promises participate in method
dispatch. \CountSThreePromiseForce promises are forced due to S3 dispatch
and \CountSFourPromiseForce are forced due to S4 dispatch.  As these numbers
are small, we ignore dispatch for the remainder of the study.

\paragraph{How often does non-local return happen?}
The \code{return} function pops the call stack until it arrives at the frame
where it originated from.  A call such as \code{f(return(1))} will,
when \code{f}'s argument is forced, return from \code{f} and its caller. So,
when a promise containing \code{return} is forced, the current function
stops executing and the stack is unwound; this is a non-local return. Only
\CountNonLocalArguments{} arguments to \CountNonLocalFunctions{} functions
performed non-local returns.  One of the most common causes is the
\code{base::tryCatch} function. The function sequentially attaches handlers
specified in its vararg list and executes \code{exp} by wrapping it in a
call to \code{return}. The return causes the control flow to exit
\code{doTryCatch} and \code{tryCatch}.

\begin{lstlisting}[style=R]
 tryCatch <- function(exp, ..., fin) {
 ...
     doTryCatch <- function(e, nm, penv, handler) {
         .Internal(.addCondHands(nm,handler), penv, environment(), F))
         e
     }
     value <- doTryCatch(return(exp), nm, penv, handler)
 ...
\end{lstlisting}
\vspace{2mm}

\paragraph{Takeways.} Promises dominate the memory  profile of R programs. 
They are short lived, 80\% are evaluated in the called function and over
99\% do not survive a single GC cycle. The vast majority of promises contain
a value or a variable. Only \PercArgumentPromiseExpressionFunctionCall
contain code that needs to be evaluated. Of those expression-carrying
promises \PercPromiseFunctionCallNonEvaluation are unused,
\PercPromiseFunctionCallCallerEvaluation are evaluated in the called
function and \PercPromiseFunctionCallRemoteEvaluation are evaluated down the
call-stack or meta-programmed. Overall most promises lead a rather mundane
life that one would hope a compiler could optimize out of existence.

\subsection{Strictness}

Our second research question concerns strictness. A function is said to be
\emph{strict} if it evaluates all of its arguments in a single pre-ordained
order (\emph{e.g.}, left to right).

\begin{center}
{\bf RQ2}: \emph{What proportion of R functions are strict?}
\end{center}

\begin{figure}[!h]\includegraphics[width=\textwidth]{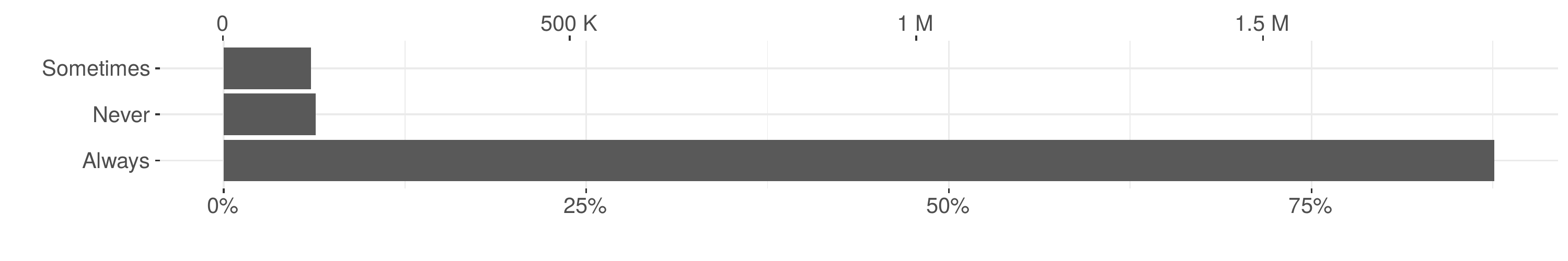}
\vspace{-8mm}\caption{Parameter strictness}\label{parstrict}
\end{figure}

\noindent
To answer this question we start with individual parameters. We only
consider plain R functions that are called more than once, have at least one
parameter and have not abruptly stopped executing owing to non-local returns.
There \CountParameter distinct parameters to such functions. For
a given parameter and a given function, we aggregate all calls and all uses
of that parameter into three categories: parameters that are
{\it\underline{Always}} evaluated, parameters that are
{\it\underline{Never}} evaluated, and {\it\underline{Sometimes}} evaluated.
Figure~\ref{parstrict} summarizes this analysis. \PercAlways{} of parameters
are always evaluated, \PercSometimes parameters are evaluated in some calls
and not others, \PercNever parameters are never used.

\begin{figure}[!h] \centering
\includegraphics[width=\textwidth]{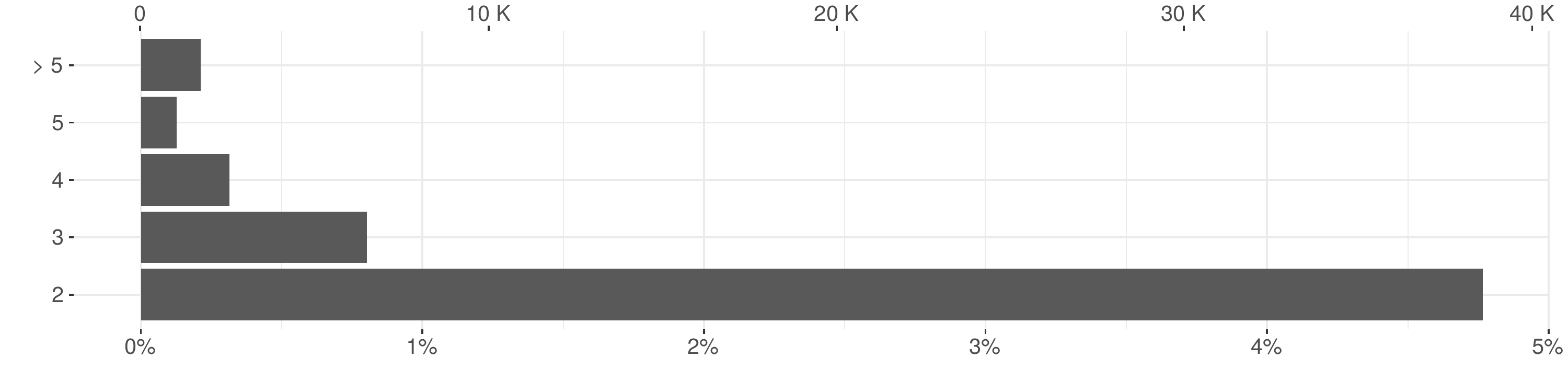}
\vspace{-1cm}
\caption{Function force orders} \label{figure:force_order}
\end{figure}

\noindent
A function is strict if its parameters are always evaluated in the same
order.  Most functions, 93.6\% to be precise, have a single order of
evaluation. This means they are candidate for being strict. Functions with
multiple orders of evaluation for their arguments are summarized in
Figure~\ref{figure:force_order}.  About 4.7\% of the functions have two
force orders, and very few functions have more.

\begin{figure}[!h]\includegraphics[width=\textwidth]{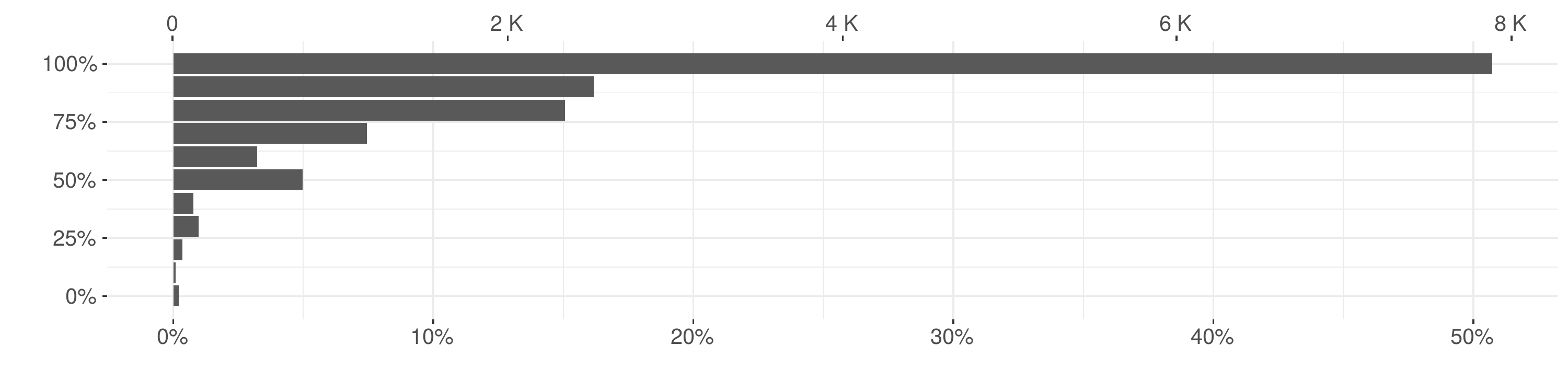}
\vspace{-8mm}
\caption{Strictness per package (x-axis=packages; y-axis=strict
  function ratio) }\label{packstrict}
\end{figure}

\vspace{-1mm}
\noindent
Based on the above data, out of a total of \CountFun functions,
\PercStrictFun are strict. Figure~\ref{packstrict} gives a histogram of
function strictness ratios per package. The majority of packages contain
only strict functions.  The packages that are less than 75\% strict account
for only \CountStrPackLess packages (\PercStrPackLess of all packages) and
\CountStrFunLess functions (\PercStrFunLess of all functions).

\begin{figure}[!h] \centering
\includegraphics[width=\textwidth]{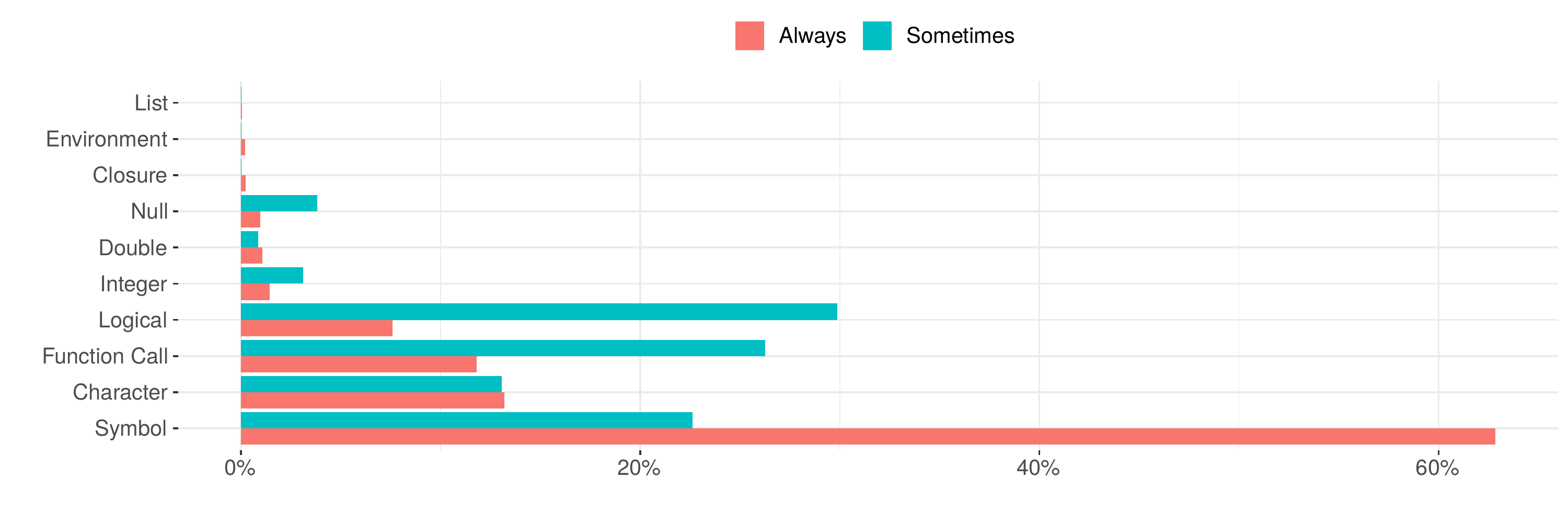}
  \vspace{-1cm}
\caption{Promise expressions} \label{figure:param_expr_types}
\end{figure}

\vspace{-2mm}
\noindent
One could consider relaxing strictness to allow multiple orders of
evaluation if those orders of evaluation could be shown to be semantically
equivalent.  Some features of R may help. First, all vectors have a
copy-on-write semantics, thus many side effects are hidden from view.
Moreover, as Figure~\ref{figure:param_expr_types} shows only 25\% of
{\underline{Sometimes}} promises perform any computation. In our experience
most of those computations are side effect free.

\vspace{-2mm}
\noindent
We performed an additional analysis to get an upper bound on the side
effects performed during promise evaluation. We considered only variable reads
and writes, external side effects such as filesystem operations were not
taken into account. A meager \CountEffectPromises
promises (out of \CountProm) perform any side-effecting computation. 
There are several cases to consider, the simplest when the promise
performs a side effect to its local environment. For example, consider
the {\tt stats::power} function:

\begin{lstlisting}[style=R]
 power <- function() {
     linkinv <- function(eta) pmax(eta^(1/lambda), .Machine$double.eps)
     mu <- linkinv(eta <- eta + offset)
 ...
\end{lstlisting}
\vspace{2mm}

\noindent
The call to \c{linkinv} takes, as argument, an expression that performs a
side effect to the local variable \c{eta}. This could have been avoided by
moving the assignment above the call to \c{linkinv} but the programmer
likely wanted to save one line. This kind of local side effect can affect
other promises coming from the same environment (which is not the case
here).  Of the side-effecting promises, \PercEffectSamePromises are local.
Non-local effects can be performed, e.g., using the \c{<<-} operator to
assign to a variable in the lexically enclosing environment.  The following
is snippet from a test script in \code{cliapp} package.  The argument to
\code{capt0} modifies \code{id} using \code{<<-}.
\begin{lstlisting}[style=R]
 test_that("auto closing", {
     id <- ""
     f <- function() capt0(id <<- cli_par(class = "xx"))
     capt0(f())
 ...
\end{lstlisting}
\vspace{2mm}

\noindent Out of the side-effecting promises, \PercEffectLexicalPromises
affect a parent environment. Finally, \PercEffectOtherPromises promises
perform side effects in other environments.  The
\code{methods::callNextMethod} function is among the most common sources of
mutation of other environments.
\begin{lstlisting}[style=R]
 callNextMethod <- function (...) {
 ...
     callEnv <- parent.frame(1)
     assign(".nextMethod", nextMethod, envir = callEnv)
 ...
\end{lstlisting}
\vspace{2mm}

\noindent
We further count how many of those side effects are performed directly by
assignments occurring textually in the promise.  Of the
\CountEffectSamePromises promises performing local side effects,
\PercEffectDirectSamePromises perform side effects directly.  Of the
\CountEffectOtherPromises promises performing side effects in other
environments, \PercEffectDirectOtherPromises perform side effects directly.
Of the \CountEffectLexicalPromises promises performing side effects in the
lexical parent environment, \PercEffectDirectLexicalPromises perform
side effects directly.

\paragraph{Qualitative analysis}
We inspected 100 randomly selected functions that our dynamic analysis
marked as strict. Out of those, 82 were indeed strict.  The remaining 18
were not, but we did not observe their laziness. The majority (16 out of 18)
of incorrectly labeled functions were not using all of their parameters,
using them along some execution paths or returning early. We found a single
function that was lazy because it called another function that was itself
lazy in that particular argument and one function for which an argument
escaped. The functions that do not evaluate all parameters could be cases
where computational effort is saved if the arguments passed are complex
expressions. We also found occurrences of explicit argument forcing.
Programmers write code such as \code{x<-x} or \code{force(x)} to ensure that
argument are values. An example of such code is the
\code{scales::viridis_pal} function; it returns a closure, but forces all of
its arguments to avoid capturing their environments:

\begin{lstlisting}[style=R]
 function(alpha=1, begin=0, end=1, dir=1) {
     force_all(alpha,begin,end,dir)
     function(n) viridis(n,alpha,begin,end,dir)
 }
\end{lstlisting}
\vspace{2mm}

\noindent
The authors of higher-order functions such as \code{apply} or \code{reduce}
often enforce strictness after receiving bug reports from end-users around
unwanted lazy evaluation and variable capture interactions. The
\code{forceAndCall} function has been introduced to mitigate this issue by
forcing the arguments of a function prior to calling it. Looking for uses of
\code{forceAndCall} revealed additional packages that enforce strictness for
higher-order functions. The function is invoked \CountForceAndCallCall
times. The \code{force} function is also widely used to enforce strictness;
it is called \CountForceCall times and used in 60\% of the packages we
inspected.  In summary, manual inspection suggests that we overestimate
strictness. Improving precision of the analysis would require increasing
code coverage. We also found numerous occasions where programmers require
strictness to control side effects.

\subsection{Meta-programming}

The next research question pertains to the use of call-by-need to enable
meta-programming.

\begin{center}
{\bf RQ3}: \emph{How frequently are promises used for meta-programming?}
\end{center}

\begin{figure}[!h]
\includegraphics[width=\textwidth]{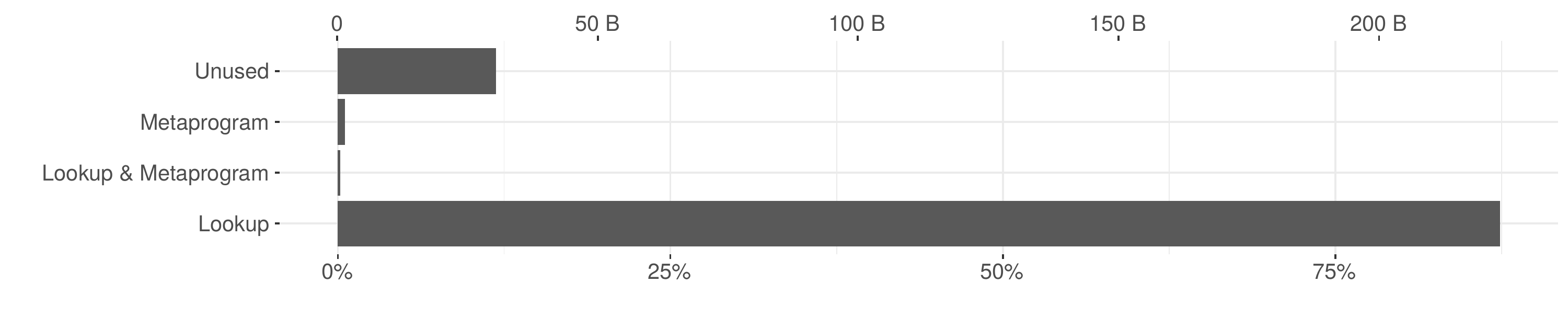}
\vspace{-8mm}
\caption{Meta-programmed promises}\label{lookmeta}
\end{figure}

\noindent
For the purposes of this discussion we define meta-programming as the
manipulation of code through calls to \code{substitute} which lets
programmers extract an abstract syntax tree from the body of a promise,
modify it, and evaluate with \code{eval}.  We observed \CountSubstituteCall
calls (\PercSubstituteCall of all calls) to \code{substitute}.
Figure~\ref{lookmeta} shows the number of promises that were
meta-programmed. The graph has four categories: promises that were created
but never used, promises that were meta-programmed, promises that were both
meta-programmed and accessed, and lastly, promises that were only accessed.
The data shows that \PercMetaOnly of promises were used purely for
meta-programming purposes, while \PercLookMeta were both forced to obtain a
value and used for meta-programming.

\noindent
Meta-programming is widespread with \CountPackMeta (\PercPackMeta) packages
using it. One feature of \code{substitute} is that programmers can specify
the environment in which to resolve names occurring in its argument.  It is
also possible to access an environment up the call chain and invoke
\code{substitute} on it.  Over \PercCallerEnvSubstituteCalls of calls to
\code{substitute} use the current environment.  Only
\PercNewEnvSubstituteCalls of calls use a custom replacement list or a new
environment, and \CountDynamicEnvSubstituteCalls use a parent frame.  These
are almost entirely due to \code{deparse}, \code{eval} and \code{do.call}
which allow specifying their arguments' evaluation environment.  For
example, the \code{envnames::get_env_names} function uses \code{substitute}
in different contexts to extract user-defined names of environments. The
first call to \code{substitute} runs in the second frame from the top of the
stack, the next call in the first frame, and the third call in the
environment in which it is called.

\begin{lstlisting}[style=R]
 get_env_names <- function(envir=NULL, include_functions=FALSE) {
     get_informative_environment_name <- function(envir)
         if (...) envir_name <- deparse(substitute(envir, parent.frame(n = 2)))
         else     envir_name <- deparse(substitute(envir, parent.frame(n = 1)))
     if (!is.null(envir) && !is.environment(envir))
         { error_NotValidEnvironment(deparse(substitute(envir)));  return(NULL) }
     envir_name <- get_informative_environment_name(envir)
 ...
\end{lstlisting}
\vspace{2mm}

\noindent
The \code{glmmTMB::makeOp} function constructs ASTs from replacement lists
instead of an environment. Arguments \code{x}, \code{y} and \code{op} are
evaluated and bound to \code{X}, \code{Y} and \code{op} respectively in the
\code{list} which is then used for replacement in the AST by
\code{substitute}.
\begin{lstlisting}[style=R]
makeOp <- function(x, y, op=NULL) {
    if (is.null(op) || missing(y)) {
        if (is.null(op)) substitute(OP(X), list(X = x, OP = y))
        else               substitute(OP(X), list(X = x, OP = op))
    } else substitute(OP(X, Y), list(X = x, OP = op, Y = y))
}
\end{lstlisting}

\paragraph{Qualitative analysis}
We manually inspected 100 functions that meta-program their arguments, and
classified them based on the usage patterns.  One common pattern is to
extract the source text of an argument. This is used by various plotting
functions to give default names to the axes of a graph if none are
provided. In the following definition \code{xAxis} and \code{yAxis} are
parameters used that way.  The call to \code{substitute} returns the AST of
the arguments, and \code{deparse} turns those into text.

\begin{lstlisting}[style=R]
 function(design, xAxis, yAxis) {
     designName <- deparse(substitute(design))
     xAxisName <- deparse(substitute(xAxis))
     yAxisName <- deparse(substitute(yAxis))
     plot(1, type = "n", main = designName, xlab = xAxisName, ylab = yAxisName,
 ...
\end{lstlisting}
\vspace{2mm}

\noindent
The following pattern explains why many promises are both evaluated and
meta-programmed. The call to \code{substitute} extracts the AST for \code{deg}
and the next line evaluates \code{deg}.

\begin{lstlisting}[style=R]
 function(deg) {
     degname <- deparse(substitute(deg))
     deg <- as.integer(deg)
     if (deg < 0 || deg > 1) stop(paste0("Error ",degname))
     deg
 }
\end{lstlisting}
\vspace{2mm}

\noindent
Another common pattern, that is a syntactic convenience, is to allow the use
of symbols instead of strings. The \code{::} operator is used to prefix
function names with their packages. It is implemented as a reflective
function that expects two strings. But programmers would rather write
\code{base::log} than \code{"base"::"log"}. To this end, the arguments are
left uninterpreted, instead the function deparses them to strings.

\vspace{2mm}
\begin{lstlisting}[style=R]
 `::` <- function(pkg, name) {
     pkg <- as.character(substitute(pkg))
     name <- as.character(substitute(name))
     get(name, envir=asNamespace(pkg), inherits=FALSE)
 }
\end{lstlisting}
\vspace{2mm}

\noindent
Meta-programming is also used for better error reporting and logging. This
example shows code that only retrieves the source text of the argument.

{\small \begin{lstlisting}[style=R]
 function(arg) 
     if (!is.numeric(arg)) stop(paste(deparse(substitute(arg)),"is not numeric"))
\end{lstlisting}}
\vspace{2mm}

\noindent
We also found functions that leverage non-standard evaluation. The following
definition is for \code{base::local} which provides limited form of
sandboxing by evaluating code in a new environment.  The argument is
extracted and evaluated using \code{eval} in an empty or user-supplied
environment.

\begin{lstlisting}[style=R]
 function(arg, envir=new.env()) eval.parent(substitute(eval(quote(arg),envir)))
\end{lstlisting}
\vspace{2mm}

\noindent
A combination of meta-programming, dynamic evaluation and first-class environments opens
up the door for domain specific languages. The pipe operator heavily used in
the \code{tidyverse} group of packages performs non-standard evaluation on
its arguments. While the user writes code like this \code{df \%>\% mean},
what is actually executed is \code{mean(df)}. While the actual definition relies
on more intricate non-standard evaluation techniques, a simple definition that
achieves a similar effect turns both arguments into abstract syntax trees, and
captures the calling environment.

\begin{lstlisting}[style=R]
 function(lhs, rhs) {
     lhs <- substitute(lhs)
     rhs <- substitute(rhs)
     eval(call(pipe, rhs, lhs), parent.frame(), parent.frame())
 }
\end{lstlisting}
\vspace{2mm}

\noindent
Overall, the use of meta-programming is widespread and falls in two rough
categories: access to the source text of an argument in the direct caller
and non-standard evaluation of an argument. The latter is the source of much
of the expressive power of the language and is critical to some of the
most widely used packages such as \code{ggplot} and \code{dplyr}.

\subsection{Revisiting the Traditional Benefits of Laziness}

The next research question asks whether the benefits of lazy evaluation that
were advocated by \citet{Hudak89} are realized in the R ecosystem.

\begin{center}
{\bf RQ4}: \emph{Are the traditional benefits of laziness realized in R?}
\end{center}

\begin{figure}[!b]
\includegraphics[width=\textwidth]{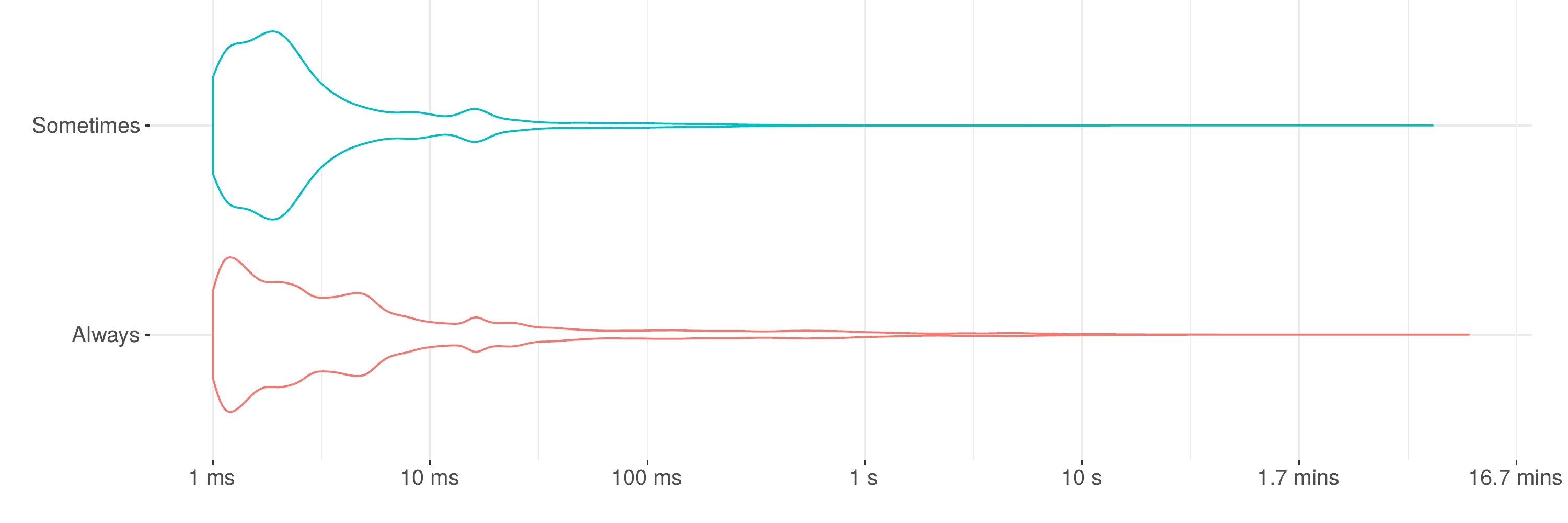}
\vspace{-8mm}
\caption{Promise evaluation}\label{run}
\end{figure}

\noindent
These benefits are that programmers need not worry about the cost of unused
arguments and they are able to define and use unbounded data structures. We
posit the following hypothesis, if programmers understand how call-by-need
works, they will feel free to pass complex computations in non-strict
positions. If true, one could hope to observe a difference in running time
for arguments known to be strict (promises passed to {\it Always}
parameters) and those that are not (promises passed to {\it Sometimes}
parameters).  Figure~\ref{run} shows the probability density of promise
running times with times smaller than a millisecond discarded.  The promises
that are passed to {\it Sometimes} arguments tend to be less expensive to
evaluate. While there is a difference in the profiles, the data does not
allow us to confirm that programmers are taking advantage of laziness.

\begin{figure}[!h]
\includegraphics[width=\textwidth]{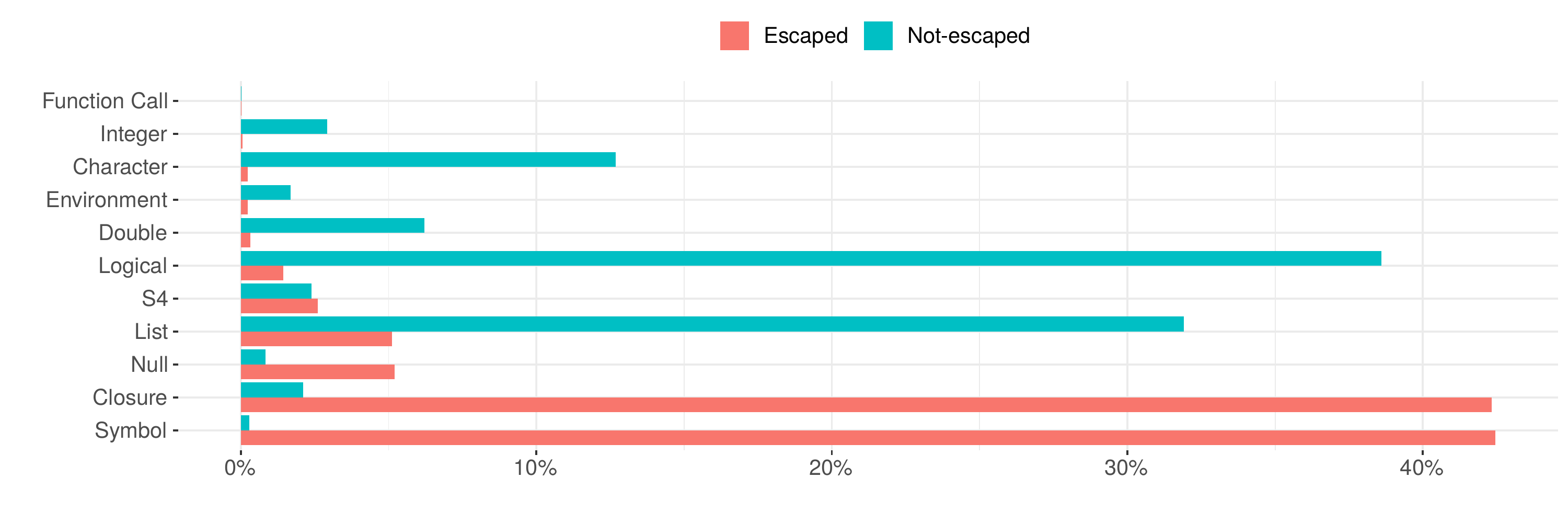}
\vspace{-8mm}
\caption{Function results}\label{escrettyp}
\end{figure}

\noindent
Laziness makes it possible to compute over infinite data structures.  While
R does not provide such data structures, it is conceivable that programmers
created some in their code.  As we have shown, one can use promises together
with environments to create unbounded data structures. While it is difficult
to measure this directly, we can measure \emph{escaped promises}. These
promises outlive the function they are passed into, and these would be a
superset of lazy data structures. Of the \CountPromiseArgument promises we
have observed, only \CountEscapeArgument escape. This is a rather small
number. We need to establish the reason why they escape.
Figure~\ref{escrettyp} compares the return types of functions which have at
least one of their promises escaping, and functions that do not have
escaping promises. The main difference between them is that functions with
escaping promises have a large number of symbols and closures as their
return values.  The next section performs a qualitative analysis to
understand how those closures are used.

\paragraph{Qualitative analysis}
It is not easy to assess, from the quantitative results alone, whether
programmers benefit from laziness. Consider a call, \code{f(a+b,c)}, and
imagine that depending on the value of \code{c}, the first argument may or
may not be evaluated. If \code{a} and \code{b} are large matrices and
\code{c} is infrequently true, a programmer aware of laziness would not
worry about the performance of this code. Without laziness, the API of the
function would likely have to change so that instead of passing the result
of the computation one would pass the individual arguments and let the
function perform the addition if needed.  In our manual inspection, we have
not found any code suggesting that programmers are concerned about the cost
of evaluating expressions, but this is most likely due to the fact that many
users are not performance sensitive.

One promising use of laziness is related to \code{delayedAssign}.  We
observed \CountDelayedAssignCall calls to this function.  We inspected
manually 36 packages that use it. We found several recurring patterns that
aim to avoid unnecessary computation. Examples are: the \code{AzureML}
package uses delayed assignment to avoid loading unneeded parts of its
workspace; \code{crunch} uses it to delay fetching data from a server; and
(slightly surprisingly) \code{callCC} uses it in conjunction with non-local
return to implement the \code{call/cc} function.  Overall we found little
evidence of programmers taking advantage of call-by-need, other than in
cases where they explicitly called \code{delayedAssign}, in the sample of
functions we inspected. We did find cases where the authors of the code
seemed to want to enforce a consistent evaluation order and prevent
argument-induced side effects from happening in the midst of evaluation of
the function.

\begin{figure}[!h]
\flushleft
{\bf Closure:}
\vspace{-0.8mm}
{\small\begin{lstlisting}[style = R]
 function(arg, e=10^-5) { function(x) (arg(x+e)-arg(x-e))/(2*e) }
\end{lstlisting}}
\vspace{0mm}
{\bf S4 object:}
\vspace{-0.8mm}
{\small\begin{lstlisting}[style = R]
 function(arg)  new("FLXcomponent",df=arg$df)
\end{lstlisting}}
\vspace{0mm}
{\bf Environment:}
\vspace{-0.8mm}
{\small\begin{lstlisting}[style = R]
 function(arg) { env<-new.env();  env$fn<-function(x){...out<-arg(x)...}; env }
\end{lstlisting}}
\vspace{0mm}
{\bf Finalizer:}
\vspace{-0.8mm}
{\small\begin{lstlisting}[style = R]
 function(arg)  reg.finalizer(environment(), function(...) dbDisconnect(arg))
\end{lstlisting}}
\vspace{0mm}
{\bf Delayed assignment:}
\vspace{-0.8mm}
\begin{lstlisting}[style = R]
 function(arg, e) delayedAssign(x, get(from, arg), assign.env=as.environment(e)))
\end{lstlisting}
\vspace{0mm}
{\bf Formula:}
\vspace{-0.8mm}
{\small\begin{lstlisting}[style = R]
 function(arg, i) as.formula(arg[, 1] ~ arg[, i])
\end{lstlisting}}
\caption{Escaping promises}\label{escape}
\end{figure}
\newpage
\noindent
To detect an infinite data structure we looked at occurrences of promises
that outlive the function in which they were passed. We inspected 100
functions with escaping arguments and observed the following patterns: (1)
arguments captured in closures, (2) arguments captured in S4 objects, (3)
arguments stored in environments, (4) arguments passed into finalizers, (7)
argument passed into delayed assignments, and (8) arguments passed into
formulas. Figure~\ref{escape} gives examples of each of these categories.
In terms of linguistic mechanisms, all but the last two end up as variants
of closure-captured promises.  Formula is interesting, because it is really
a domain specific language that is interpreted with different semantics.  In
our time spent working with R, we found a single package,
\code{Rstackdeque}~\cite{oneil15} that advertised the use of lazy data
structures, specifically fully persistent queues based on \citep{oka95}.
This package, which depends on lazy lists, is the only use of lazy data
structures we are aware of in the R ecosystem.

\section{Related Work}

Lazy functional programming languages have a rich history. The earliest lazy
programming language was Algol 60~\cite{Backus} which had a call-by-name
evaluation strategy. This was followed by a series of purely functional lazy
languages~\cite{Turner79,Augustsson93,Turner85}. The motivations for the
pursuit of laziness were modularity, referential transparency and the
ability to work with infinite data structures~\cite{hughes}.  These
languages inspired the design of Haskell~\cite{haskell}.

The meta-programming support of R is reminiscent of fexprs~\cite{mitch} in
Lisp. Fexprs are first class functions with unevaluated arguments.  
In R, functions always have access to their unevaluated and evaluated
arguments.  Pitman~\cite{Pitman80} argued in favor of macros over fexprs.
Macros are transparent, their definition can be understood by expanding them
to primitive language forms before the evaluation phase. fexprs on the other
hand perform code manipulation during evaluation. This makes it harder for
compilers to statically optimize fexprs. Furthermore, expression
manipulation such as substitution of an expression for all evaluable
occurrences of some other expression can be performed correctly by macros
because they expand before evaluation to primitive forms.

Purdue FastR~\cite{vee14} is an AST interpreter for R written in Java to
explore the applicability of simple compiler optimization techniques, within
the reach of scientific community lacking expertise in language
run-times. The authors implement an optimization technique that defers
element wise operations on vectors by constructing expression trees called
Views, which are evaluated on demand. This prevents the materialization of
temporary vectors in a chain of vectorized mathematical operations. Like
promises, views cache the result of evaluating the expression. However,
unlike promises which are exposed to the user through meta-programming,
views are completely transparent to the user. Promises are built by
packaging arbitrary argument expressions but views are built incrementally
by piling referentially transparent vector operations such as \code{+},
\code{-}, \code{log}, \code{ceil}, etc. Promises are evaluated very quickly
due to the eager nature of most functions, but the expression trees of views
are evaluated only when the entire result vector or its subset is demanded
or a selected aggregate operation such as \code{sum} is applied.

Building upon the implicit argument quoting of promises is a data structure
called quosure, short for quoted closure, that bundles an expression and its
evaluation environment for explicit manipulation at the the language
level. A quosure is thus an explicit promise object exposed to the user,
with APIs to access the underlying expression and environment. Quosures are
a central component of a collection of R packages for data manipulation,
Tidyverse~\cite{tidyverse}, that have a common design language and
underlying data structures. Dplyr~\cite{dplyr}, a package of Tidyverse,
implements a DSL for performing SQL like data transformations on tabular
data and ggplot2~\cite{ggplot2} implements a declarative language for
graphing data, inspired by the Grammar of Graphics. These packages quote,
unquote and quasiquote user supplied expressions and evaluate them in
appropriate environments. To facilitate this, these packages also provide an
evaluation function, \code{eval_tidy} that extends the base \code{eval} to
deal with quosures. This suggests that reifying promises can be useful.

Renjin is an implementation of R built on the Java virtual machine designed
to analyze large data sets and facilitate integration with enterprise
systems.  Renjin supports delaying evaluation of side effect free
computation~\cite{alex}. Instead of returning the actual result of a
computation, Renjin returns placeholders which look and behave exactly like
the actual computation result, but will only calculate results if forced to.
The difference between Renjin and FastR, both systems are more ``lazy'' than
GNU R, lies in Renjin's support for relational-style optimization.

\citet{ecoop12} implemented a tool called TraceR for profiling R programs.
The architecture of TraceR was similar to that of the pipeline presented
here, but it did not target large scale data collection and has gone
unmaintained for several years. Our current infrastructure is less invasive
than the previous implementation and is being considered for inclusion in
\GNUR.

Finally, we compare our semantics to the work of \citet{trouduq}.  Our
semantics makes no claims of being correct (there is no specification of R)
or of being faithful to the language. The semantics is useful in as much it
provides a readable account of delayed evaluation in R. Bodin's work is more
ambitious, it aims to provide an executable semantics.  The benefits of
executable semantics is that they can be tested against an implementation,
in this case the GNU R virtual machine.  The semantics consists of 28,026
lines of Coq and 1,689 lines of ML.  Validation is done through testing and
visual comparison between the GNU R's C code and Coq code.
Unfortunately, in the current state Bodin's specification is still far
from complete. Out of 20,976 tests, only 6,370 pass.  Inspection of the
specification reveals that key functions for laziness such as \code{force},
\code{forceAndCall}, and \code{delayedAssign} are not implemented. Only a
handful of the provided tests deal with lazy evaluation (they check that
promises are evaluated only when forced). Furthermore, package loading and
interaction with C code is not supported, thus packages from our corpus
cannot be tested.  We tried to match our semantics to theirs but the DLS'18
paper does not describe their treatment of laziness. Due to the size of the
Coq codebase and lack of documentation, it was unclear how to align the two
artifacts.

\newpage
\section{Conclusion}

This paper offers a glimpse into the design, implementation and usage of
call-by-need in the R programming language. Call-by-need is the default in
R, but our data suggests that it is used less than one would expect.  To
deal with side effects and manage programmers' expectation, many functions
are stricter than they need to be.  We found little evidence of lazy data
structures or that users leverage lazy evaluation to avoid unnecessary
computation.  We found only two broad categories of usages that benefited
from it. The first is the creation of delayed bindings. These, in our
experience, are always explicit. The second is for meta-programming. Within
that category, uses are split between accessing the source text of an
expression for debugging purposes and performing non-standard evaluation.

The costs of lazy evaluation in performance and memory use are substantial.
Every argument to a function must be boxed in a promise, retaining a
reference to the function's environment until evaluated. Every access to a
variable must check if it is bound to a promise and either evaluate it or
read the cached value. Lazy evaluation complicates the task of compilers and
program analysis tools as they must deal with the possibility of any
variable access causing side effects. Lastly, the majority of users do not
expect arguments to be evaluated in a lazy fashion, thus leading to hard to
understand bugs.

If laziness is mostly unused, could it be eliminated?  Any change to the
semantics of a widely used language has to be minimally invasive. We are
considering the following combination of ideas. For the meta-programming
use-cases that require only source code we propose to offer a function which
returns the caller expression for any argument, this is possible as the
information is present in the debug meta-data of the interpreter. For
functions that do non-standard evaluation, we propose to add annotations on
their parameters to request a promise to be generated. This can be implemented by
adding a run-time check before function calls.

While we propose removing laziness, there is also an argument for
strengthening it. In many ways, R is only \emph{weakly} lazy, it forces
promises in many places where other languages would not. The works of
\citet{tidyverse}, \citet{alex} and \citet{vee14} suggest that more
laziness can bring interesting optimization opportunities, especially when
performing operation on large data objects.

\vspace{3cm}

\section*{Acknowledgments}
We thank the reviewers for constructive comments that helped us improve the
presentation. Early prototypes of our analysis were implemented by Konrad
Siek and Jan Noha, we thank them for their contribution. We would also like
to thank Ben Chung, Artem Pelenitsyn, Filip K\v{r}ikava, Tomas Kalibera, Luke
Tierney, Stepan Sindelar, Alex Bertram, and Hadley Wickham for their
comments and encouragement.  This work received funding from the Office of
Naval Research (ONR) award 503353, the European Research Council under the
European Union's Horizon 2020 research and innovation programme (grant
agreement 695412), the NSF (awards 1518844, 1544542, and 1617892), and the
Czech Ministry of Education, Youth and Sports (grant agreement
CZ.02.1.01\/0.0\/0.0\/15\_003\/0000421).

\newpage



\end{document}